\pdfoutput=1
\documentclass[format=acmsmall, review=false, screen=true]{acmart}

\usepackage{bm}
\usepackage{booktabs} 
\usepackage{pgfplots}
\usepackage{xspace}
\usepackage{subfig}
\usepackage{chronosys}
\usepackage{enumerate}
\usepackage[inline]{enumitem}
\usepackage{tabu}
\usepackage{multirow}
\usepackage[ruled]{algorithm2e}
\usepackage{hyperref}           
\hypersetup{
	colorlinks=true,
	linkcolor=blue,
	filecolor=red,      
	urlcolor=magenta,
	breaklinks=true,            
}
\usepackage{breakurl}           
\pgfplotsset{compat=1.13}
\usetikzlibrary{arrows,positioning,decorations,shapes}

\SetAlFnt{\small}
\SetAlCapFnt{\small}
\SetAlCapNameFnt{\small}
\SetAlCapHSkip{0pt}
\IncMargin{-\parindent}

\acmJournal{TKDD}
\acmArticle{1}
\acmYear{2018}

\setcopyright{none}

\begin{document}
\title[Continuous-Time Relationship Prediction]{Continuous-Time Relationship Prediction in Dynamic Heterogeneous Information Networks}

\author{Sina Sajadmanesh}
\orcid{0000-0002-8834-0338}
\affiliation{%
	\institution{Department of Computer Engineering, Sharif University of Technology}
	\streetaddress{Azadi Ave}
	\city{Tehran}
	\state{Tehran}
	\postcode{1458889694}
	\country{Iran}}
\email{sajadmanesh@ce.sharif.edu}

\author{Sogol Bazargani}
\affiliation{%
	\institution{Department of Computer Engineering, Sharif University of Technology}
	\streetaddress{Azadi Ave}
	\city{Tehran}
	\state{Tehran}
	\postcode{1458889694}
	\country{Iran}}

\author{Jiawei Zhang}
\affiliation{%
	\institution{IFM Lab, Department of Computer Science, Florida State University}
	\streetaddress{1017 Academic Way}
	\city{Tallahassee}
	\state{Florida}
	\postcode{32304}
	\country{United States}}
\email{jiawei@ifmlab.org}

\author{Hamid R. Rabiee}
\affiliation{%
	\institution{Department of Computer Engineering, Sharif University of Technology}
	\streetaddress{Azadi Ave}
	\city{Tehran}
	\state{Tehran}
	\postcode{1458889694}
	\country{Iran}}
\email{rabiee@sharif.edu}

\renewcommand{\shortauthors}{Sajadmanesh et al.}
\newcommand{\descr}[1]{\smallskip\noindent\textbf{#1}}
\newcommand{\npglm}{{\textsc{Np-Glm}}\xspace}
\newcommand{\mb}[1]{\mathbf{#1}}
\newcommand{\mc}[1]{\mathcal{#1}}
\newcounter{sarrow}
\newcommand\xrsquigarrow[1]{%
	\stepcounter{sarrow}%
	\begin{tikzpicture}[decoration=snake]
	\node (\thesarrow) {\strut#1};
	\draw[->,decorate] (\thesarrow.south west) -- (\thesarrow.south east);
	\end{tikzpicture}%
}

\begin{abstract}
Online social networks, World Wide Web, media and technological networks, and other types of so-called \emph{information networks} are ubiquitous nowadays. These information networks are inherently \emph{heterogeneous} and \emph{dynamic}. They are heterogeneous as they consist of multi-typed objects and relations, and they are dynamic as they are constantly evolving over time. One of the challenging issues in such heterogeneous and dynamic environments is to forecast those relationships in the network that will appear in the future. In this paper, we try to solve the problem of continuous-time relationship prediction in dynamic and heterogeneous information networks. This implies predicting the time it takes for a relationship to appear in the future, given its features that have been extracted by considering both heterogeneity and temporal dynamics of the underlying network. To this end, we first introduce a feature extraction framework that combines the power of meta-path-based modeling and recurrent neural networks to effectively extract features suitable for relationship prediction regarding heterogeneity and dynamicity of the networks. Next, we propose a supervised non-parametric approach, called \emph{Non-Parametric Generalized Linear Model} (\npglm), which infers the hidden underlying probability distribution of the relationship building time given its features. We then present a learning algorithm to train \npglm and an inference method to answer time-related queries. Extensive experiments conducted on synthetic data and three real-world datasets, namely Delicious, MovieLens, and DBLP, demonstrate the effectiveness of \npglm in solving continuous-time relationship prediction problem vis-\`a-vis competitive baselines. 
\end{abstract}

%
%

\ccsdesc[500]{Information systems~Data mining}
\ccsdesc[500]{Information systems~Social recommendation}
\ccsdesc[300]{Computing methodologies~Machine learning}

%
%

\keywords{Link Prediction, Social Network Analysis, Heterogeneous Network, Non-Parametric Modeling, Recurrent Neural Network, Autoencoder}

\maketitle

\section{Introduction}\label{sec:intro}
Link prediction is the problem of prognosticating a certain relationship, like interaction or collaboration, between two entities in a networked system that are not connected already \cite{lu2011link}. Due to the popularity and ubiquity of networked systems in the real world, such as social, economic, or biological networks, this problem has attracted a considerable attention in recent years and has found its applications in various interdisciplinary domains, such as viral marketing, bioinformatics, recommender systems, and social network analysis \cite{wasserman1994social}. For example, suggesting new friends in an online social network \cite{liben2007link} or predicting drug-target interactions in a biological network \cite{chen2012drug} are two quite different problems, but can both cast as the prediction task of friendship links and drug-target links, respectively.

The problem of link prediction has a long literature and is studied extensively in the last decade. Initial works on link prediction problem mostly concentrated on homogeneous networks, which are composed of single type of nodes connected by links of the same type \cite{liben2007link, wang2007local, lichtenwalter2010new}. However, many of today's networks, such as online social networks or bibliographic networks, are inherently \emph{heterogeneous}, in which multiple types of nodes are interconnected using multiple types of links \cite{taskar2004link, shi2017survey}. For example, a bibliographic network may contain author, paper, venue, etc. as different node types; and write, publish, cite, and so on as diverse link types that bind nodes with different types to each other. In these heterogeneous networks, the concept of a link can be generalized to a relationship, which can be constructed by combining different links with different types. For instance, the author-cite-paper relationship can be defined in a bibliographic network as a combination of author-write-paper and paper-cite-paper links. Analogously, one can generalize the link prediction to \emph{relationship prediction} in heterogeneous networks which tries to predict complex relationships instead of links \cite{sun2012will}.

While most of the studies on the link/relationship prediction in heterogeneous networks utilize a static snapshot of the underlying network, many of these networks are \emph{dynamic} in nature, which means that new nodes and linkages are continually added to the network, and some existing nodes and links may be removed from the network over time. For example, in online social networks, such as Facebook, new users are joining in the network every day, and new friendship links are being added to the network gradually. This dynamic characteristic causes the structure of the network to change and evolve over time, and taking these changes into account can significantly boost the quality of link prediction task \cite{potgieter2009temporality}.

In recent years, newer studies have shifted from traditional link prediction on static and homogeneous networks toward newer domains, considering heterogeneity and dynamicity of networks \cite{dong2012link, davis2011multi, 7752228, hajibagheri2016leveraging, moradabadi2017novel}. However, most of these works merely focus on one of these aspects, disregarding the other. Although there are quite a few studies that address both the challenges of heterogeneity and dynamicity \cite{aggarwal2012dynamic, sett2017temporal}, to the best of our knowledge, all of them have ultimately formulated the link prediction problem as a binary classification task, i.e., predicting \emph{whether} a link will appear in the network in the future. However, in dynamic networks, new links are continually appearing over time. So a much more interesting problem, which we call it \emph{continuous-time link prediction} in this paper, is to predict \emph{when} a link will emerge or appear between two nodes in the network. Examples of this problem include predicting the time at which two individuals become friends in a social network or the time when two authors collaborate on writing a paper in a bibliographic network \cite{sun2012will}. Inferring the link formation time in advance can be very useful in many concrete applications {in different disciplines, such as socialogy, economics, biology, and epidemiology, where the interactions between entities can be modeled via timed links}. For example in the biological context, predicting the marker proteins interaction time in a gene regulatory network will lead to predicting tumor progression and prognosis \cite{taylor2009dynamic}. As another example in online social networks, if the recommender system could predict the relationship building time between two people, then it can issue a friendship suggestion close to that time since it will have a relatively higher chance to be accepted. Good continuous-time link prediction results will lead to denser connections among users, and can greatly improve users' engagement that is the ultimate goal of online social networks \cite{kwak2010twitter}.

In this paper, we aim to solve the problem of continuous-time relationship prediction, in which we forecast the relationship building time between two nodes in a dynamic and heterogeneous environment. This problem is very challenging from the technical perspective, and cannot be solved trivially for three main reasons. First, the formulation of continuous-time relationship prediction is quite different from the conventional link prediction due to the involvement of temporal dynamics of the network and the necessity of considering network evolution time-line. Second, we only know the building time of those relationships that are already present at the network and for the rest of them that are yet to happen, which are excessive in number versus the existing ones, we lack such information. Finally, as opposed to the works concerning the binary link prediction, there are very rare works in the literature on continuous-time link prediction that attempt to answer the ``when'' question. To the best of our knowledge, the only work that has studied the continuous-time relationship prediction problem so far is proposed by Sun et al. \cite{sun2012will}. They infer a probability distribution over time for each pair of nodes given their features and answer time-related queries about the relationship building time between the two nodes using the inferred distribution. However, the drawback of their method, not to mention neglecting the temporal dynamics of the network, is that it mainly relies on the assumption that relationship building times are coming from a certain probability distribution that must be fixed beforehand. This assumption though simplifying is very restrictive, because in real applications this distribution is unknown, and considering any specific one as a priori could be far from reality or limit the solution generality.

In order to address the above challenges, we propose a supervised non-parametric method to solve the problem of continuous-time relationship prediction. To this end, we first formally define the continuous-time relationship prediction problem and formulate the approach to solve it generally. Then, we introduce our novel feature extraction framework which leverages meta-path-based modeling and recurrent neural networks to deal with heterogeneity and dynamicity of information networks. Next, we present \emph{Non-Parametric Generalized Linear Model} (\npglm) which models the distribution of relationship building time given the extracted features. The strength of this non-parametric model is that it is capable of learning the underlying distribution of the relationship building time, as well as the contribution of each extracted feature in the network. Afterward, we propose an inference algorithm to answer queries, like the most probable time by which a relationship will appear between two nodes or the probability of relationship creation between them during a specific period. Finally, we conduct comprehensive experiments over a synthetic dataset to verify the correctness of \npglm's learning algorithm, and on three real-world dataset - DBLP, Delicious, and MovieLens - to demonstrate the effectiveness and generality of the proposed method in predicting the relationship building time versus the relevant baselines. As a summary, we can enumerate our major contributions as follows:

\begin{enumerate}[label=(\roman*)]
\item The proposed feature extraction framework can tackle heterogeneity of the data as well as capturing the temporal dynamics of the network by incorporating meta-path-based features into a recurrent neural network based autoencoder.
\item Our non-parametric model takes a unique approach toward learning the underlying distribution of relationship building time without imposing any significant assumptions on the problem.
\item Extensive evaluations over both synthetic and real-world datasets are performed to investigate the effectiveness of the proposed method. 
\item To the best of our knowledge, this paper is the first one which studies the continuous-time relationship prediction problem in both dynamic and heterogeneous network configurations.
\end{enumerate}

The rest of this paper is organized as follows. In Section \ref{sec:problem}, we provide introductory backgrounds on the concept and formally define the problem of continuous-time relationship prediction. Then in Section \ref{sec:features}, we introduce our novel feature extraction framework. Next, we go through the details of the proposed \npglm method in Section \ref{sec:method}, explaining its learning method and how it answers inference queries. Experiments on synthetic data and real-world datasets are described in Section \ref{sec:synthetic} and \ref{sec:results}, respectively. Section \ref{sec:related} discusses the related works, and finally in Section \ref{sec:conclusion}, we conclude the paper.

\begin{figure*}
    \centering
    \scriptsize
    \tikzstyle{block} = [ellipse,draw=black]
    \tikzstyle{arrow} = [thick,->,>=stealth]
    \tikzstyle{label} = [fill=white,inner sep=0,xshift=0.1cm,yshift=.03cm]
    \tikzstyle{self} = [out=-110,in=-70,loop,shorten >=1pt]
    \subfloat[DBLP\label{fig:schema:dblp}]{
        \begin{tikzpicture}
        \node[block] (P) at (0,0) {$\underline{P}aper$};
        \node[block] (V) at (0,1.25) {$\underline{V}enue$};
        \node[block] (T) at (-2.4,0) {$\underline{T}erm$};
        \node[block] (A) at (2.4,0) {$\underline{A}uthor$};
        \node(hidden) [draw=none] at (0,-1.5){};
        
        \draw [arrow] (A) -- node[label] {write}   (P);
        \draw [arrow] (V) -- node[label] {publish} (P);           
        \draw [arrow] (P) -- node[label] {mention} (T);
        \draw [arrow] (P) to [self] node[label,yshift=-.2cm] {cite} (P);
        
        \end{tikzpicture}
    }
    \hfil
    \subfloat[Delicious\label{fig:schema:delicious}]{
        \begin{tikzpicture}
        
        \node(B) [block] at (0,0) {$\underline{B}ookmark$};
        \node(T) [block] at (-2.4,0) {$\underline{T}ag$};
        \node(U) [block] at (2.4,0) {$\underline{U}ser$};
        \node(hidden) [draw=none] at (0,-1.5){};
        
        \draw [arrow] (U) -- node[label]{post} (B);
        \draw [arrow] (B) -- node[label]{has-tag} (T);
        \draw [arrow] (U) to [self] node[label,yshift=-.2cm]{contact} (U);
        
        \end{tikzpicture}
    }
    \hfil
    \subfloat[MovieLens\label{fig:schema:movielens}]{
        \begin{tikzpicture}
        
        \node[block] (M) at (0,0) {$\underline{M}ovie$};
        \node[block] (U) at (2.2,0) {$\underline{U}ser$};
        \node[block] (C) at (-1.25,1.25) {$\underline{C}ountry$};
        \node[block] (T) at (-2.2,0) {$\underline{T}ag$};
        \node[block] (G) at (1.25,1.25) {$\underline{G}enre$};
        \node[block] (A) at (1.25,-1.25) {$\underline{A}ctor$};
        \node[block] (D) at (-1.25,-1.25) {$\underline{D}irector$};
        \node(hidden) [draw=none] at (0,-1.5){};
        
        \draw [arrow] (U) -- node[label] {rate} (M);
        \draw [arrow] (M) -- node[label] {has-tag} (T);
        \draw [arrow] (M) -- node[label] {has-genre} (G);
        \draw [arrow] (A) -- node[label] {play-in} (M);
        \draw [arrow] (D) -- node[label] {direct} (M);
        \draw [arrow] (M) -- node[label] {produced-in} (C);
        
        \end{tikzpicture}
    }
    \caption{Schema of three different heterogeneous networks. Underlined characters are used as abbreviations for corresponding node types. }
    \label{fig:schema}
\end{figure*}
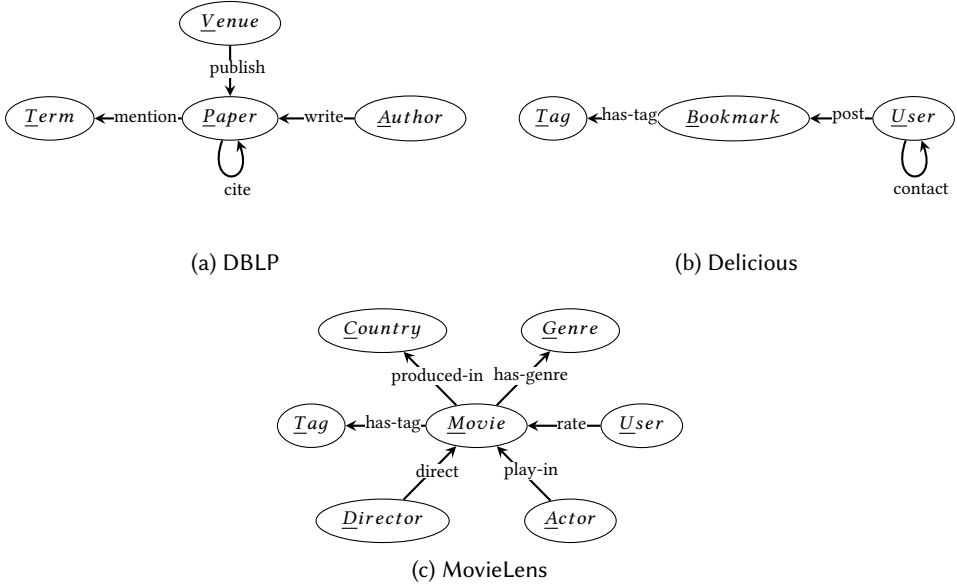

\section{Problem Formulation}\label{sec:problem}
In this section, we introduce some important concepts and definitions used throughout the paper and formally define the problem of continuous-time relationship prediction.

\subsection{Heterogeneous Information Networks}

An information network is \emph{heterogeneous} if it contains multiple kinds of nodes and links. Formally, it is defined as a directed graph $G=(V,E)$ where $V = \bigcup_i V_i$ is the set of nodes comprising the union of all the node sets $V_i$ of type $i$. Similarly, $E=\bigcup_j E_j$ is the set of links constituted by the union of all the link sets $E_j$ of type $j$. Now we bring the definition of the \emph{network schema} \cite{sun2011pathsim} which is used to describe a heterogeneous information network at a meta-level:

\begin{definition}{(Network Schema)}
    The schema of a heterogeneous network $G$ is a graph $\mc{S}_G=(\mc{V}, \mc{E})$ where $\mc{V}$ is the set of different node types and $\mc{E}$ is the set of different link types in $G$.
\end{definition}

In this paper, we focus on three different heterogeneous and dynamic networks: (1) DBLP bibliographic network\footnote{http://dblp.uni-trier.de/}; (2) Delicious bookmarking network\footnote{http://delicious.com/}; and (3) MovieLens recommendation network\footnote{https://movielens.org/}. The schema of these networks is depicted in Fig.~\ref{fig:schema}. As an example, in the bibliographic network, $\mc{V}=\left\lbrace {Author}, {Paper}, {Venue}, {Term}\right\rbrace$ is the set of different node types, and $\mc{E}=\left\lbrace\text{write}, \text{publish}, \text{mention}, \text{cite}\right\rbrace$ is the set of different link types.

Analogous to homogeneous networks where an adjacency matrix is used to represent whether pairs of nodes are linked to each other or not, in heterogeneous networks, we define \emph{Heterogeneous Adjacency Matrices} to represent the connectivity of nodes of different types:

\begin{definition}{(Heterogeneous Adjacency Matrix)}
Given a heterogeneous network $G$ with schema $\mc{S}_G=(\mc{V}, \mc{E})$, for each link type $\varepsilon\in\mc{E}$ denoting the relation between node types $\nu_i,\nu_j\in\mc{V}$, the heterogeneous adjacency matrix $M_{\varepsilon}$ is a binary $|V_{\nu_i}| \times |V_{\nu_j}|$ matrix representing whether nodes of type $\nu_i$ are in relation with nodes of type $\nu_j$ with link type $\varepsilon$ or not.
\end{definition}

For instance, in the bibliographic network, the heterogeneous adjacency matrix $M_{write}$ is a binary matrix where each row is associated with an author and each column is associated with a paper, and $M_{write}{(i,j)}$ indicates if the author $i$ has written the paper $j$.

As we mentioned in the Introduction section about heterogeneous networks, the concept of a link can be generalized to a relationship. In this case, a relationship could be either a single link or a composite relation constituted by the concatenation of multiple links that together have a particular semantic meaning. For example, the co-authorship relation in the bibliographic network with the schema shown in Fig.~\ref{fig:schema:dblp}, can be defined as the combination of two \emph{Author-{write}-Paper} links, making \emph{Author-{write}-Paper-{write}-Author} relation. When dealing with link or relationship prediction in heterogeneous networks, we must exactly specify what kind of link or relationship we are going to predict. This specific relation to be predicted is called the \emph{Target Relation} \cite{sun2012will}. For example, in DBLP bibliographic network we aim to predict if and when an author will cite a paper from another author. Thus the target relation, in this case, would be \emph{Author-{write}-Paper-{cite}-Paper-{write}-Author}.

\subsection{Dynamic Information Networks}
An information network is \emph{dynamic} when its nodes and linkage structure can change over time. That is, in a dynamic information network, all nodes and links are associated with a birth and death time. More formally, a dynamic network at the timestamp $\tau$ is defined as $G^{\tau}=(V^{\tau}, E^{\tau})$ where $V^{\tau}$ and $E^{\tau}$ are respectively the set of nodes and the set of links existing in the network at the timestamp $\tau$.

In this paper, we consider the case that an information network is both dynamic and heterogeneous. This means that all network entities are associated with a type, and can possibly have birth and death times, regardless of their types. The bibliographic network is an example of both dynamic and heterogeneous one. Whenever a new paper is published, a new \emph{Paper} node will be added to the network, alongside with the corresponding new \emph{Author}, \emph{Term}, and \emph{Venue} nodes (if they don't exist yet). New links will be formed among these newly added nodes to indicate the \textit{write}, \textit{publish} and \textit{mention} relationships. Some linkages might also form between the existing nodes and the new ones, like new \textit{cite} links connecting the new paper with the existing papers in its reference list.

In order to formally describe the state of a heterogeneous and dynamic network at any timestamp $\tau$, we define the \emph{time-aware heterogeneous adjacency matrix} in the following.
    
    \begin{definition}{(Time-Aware Heterogeneous Adjacency Matrix)}
        Given a dynamic heterogeneous network $G^\tau$ with schema $\mc{S}_G=(\mc{V}, \mc{E})$, for each link type $\varepsilon\in\mc{E}$ denoting the relation between node types $\nu_i,\nu_j\in\mc{V}$, the time-aware heterogeneous adjacency matrix $M^\tau_{\varepsilon}$ is a binary matrix representing if nodes of type $\nu_i$ are in relation with nodes of type $\nu_j$ with link type $\varepsilon$ at the timestamp $\tau$. More formally, for $a\in\nu_i$ and $b\in\nu_j$ we have:
        \[M^\tau_{\varepsilon}(a,b)=\begin{cases} 
        1, & \text{if}\quad(a,b)\in\varepsilon\quad\text{and}\quad bt(a,b) < \tau \le dt(a,b) \\
        0, & \text{otherwise}
        \end{cases}
        \]
        where $bt(a,b)$ and $dt(a,b)$ denote the birth and the death time of the link $(a,b)$, respectively.
    \end{definition}

\subsection{Continuous-Time Relationship Prediction}
Suppose that we are given a dynamic and heterogeneous information network as $G^{\tau}$ lastly observed at the timestamp $\tau$, together with its network schema $S_G$. Now, given the target relation $R$, the aim of continuous-time relationship prediction is to forecast the building time $t\ge \tau$ of the target relation $R$ between any node pair $(a,b)$ in $G^{\tau}$.

In order to solve this problem given a pair of nodes like $(a,b)$, we try to train a supervised model that can predict a point estimate on the time it takes for the relationship of type $R$ to be formed between them. The input to such a model will be a feature vector $\mb{x}$ corresponding to the node pair $(a,b)$. The model will then output with a continuous variable $t$ that indicates when the relationship of type $R$ will be built between $a$ and $b$. To train such a model, we need to assemble a dataset comprising the feature vectors of all the node pairs between which the relation $R$ have already been formed. The process of selecting sample node pairs, extracting their feature vector, and training the supervised model are explained in the subsequent sections. 


\section{Feature Extraction Framework}\label{sec:features}

In this section, we present our feature extraction framework that is designed to have three major characteristics: First, it effectively considers different type of nodes and links available in a heterogeneous information network and regards their impact on the building time of the target relationship. Second, it takes the temporal dynamics of the network into account and leverages the network evolution history instead of simply aggregating it into a single snapshot. Finally, the extracted features are suitable for not only the link prediction problem but also the generalized \emph{relationship prediction}. We will incorporate these features in the proposed non-parametric model in Section~\ref{sec:method} to solve the continuous-time relationship prediction problem.

\begin{figure}
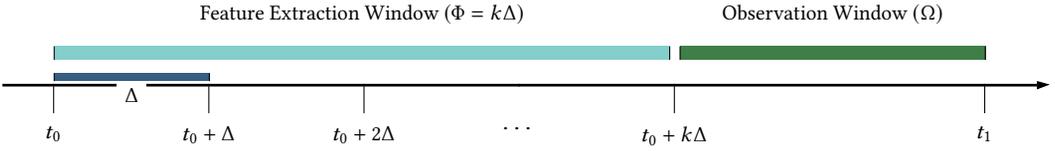

    \definecolor{blue}{HTML}{84CECC}
    \definecolor{darkblue}{HTML}{375D81}
    \definecolor{green}{HTML}{3F7F47}
    \begin{chronology}[align=left, startyear=0,stopyear=200, width=\columnwidth, height=1pt, startdate=false, stopdate=false, arrowwidth=4pt, arrowheight=3pt]
        \footnotesize
        \chronoevent[date=false]{10}{$t_0$}
        \chronoevent[date=false]{40}{$t_0+\Delta$}
        \chronoevent[date=false]{70}{$t_0+2\Delta$}
        \chronoevent[date=false,mark=false]{100}{$\dots$}
        \chronoevent[date=false]{130}{$t_0+k\Delta$}
        \chronoevent[date=false]{190}{$t_1$}
        \chronoperiode[color=darkblue, startdate=false, bottomdepth=2pt, topheight=5pt, textdepth=8pt, stopdate=false]{10}{40}{$\Delta$}
        \chronoperiode[color=blue, startdate=false, bottomdepth=10pt, topheight=15pt, textdepth=-15pt, stopdate=false]{10}{129}{Feature Extraction Window $(\Phi=k\Delta)$}
        \chronoperiode[color=green, startdate=false, bottomdepth=10pt, topheight=15pt, textdepth=-15pt, stopdate=false]{131}{190}{Observation Window $(\Omega)$}
    \end{chronology}
    \caption{The evolutionary timeline of the network data.}
    \label{fig:timeline}
\end{figure}

\subsection{Data Preparation For Feature Extraction}
To solve the problem of continuous-time relationship prediction in dynamic networks, we need to pay attention to the temporal history of the network data from two different points of view. First, we have to mind the evolutionary history of the network for feature extraction, so that the extracted features reflect the changes made in the network over time. Second, we have to specify the exact relationship building time for each pair of nodes that have formed the target relationship. This is because our goal is to train a supervised model to predict a continuous variable, which in this case is the building time of the target relationship. Hence, for each sample pair of nodes, we need a feature vector $\mb{x}$, associated with a target variable $t$ that indicates the building time of the target relationship between them.

Suppose that we have observed a dynamic network $G^{\tau}$ recorded in the interval $t_0 <\tau\le t_1$. 
According to Fig.~\ref{fig:timeline}, we split this interval into two parts: the first part for extracting the feature $\mb{x}$, and the second for determining the target variable $t$. We refer to the first interval as \emph{Feature Extraction Window} whose length is denoted by $\Phi$, and the second as \emph{Observation Window}, whose length is denoted by $\Omega$. Now, based on the existence of the target relationship in the observation window, all the node pairs in the network will fall within either one of the following three different groups:

\begin{enumerate}
    \item Node pairs that form the target relationship before the beginning of the observation window (in the feature extraction window).
    \item Node pairs that form the target relationship in the observation window for the first time (not existing before in the feature extraction window).
    \item Node pairs that do not form the target relationship (neither in the feature extraction window nor in the observation window).
\end{enumerate}

The node pairs in the 2nd and 3rd categories constitute our data samples, and will be used in the learning procedure to train the supervised model. For such pairs, we extract their feature vector $\mb{x}$ using the history available in the feature extraction window. For each node pair in the 2nd category, we see that the target relationship between them has been created at a time like $t_r\in(t_0+\Phi,t_1]$. So we set $t=t_r-(t_0+\Phi)$ as the time it takes for the relationship to form since the beginning of the observation window. For these samples, we also set an auxiliary variable $y=1$ which indicates that we have \emph{observed} their exact building time. On the other hand, For node pairs in the 3rd category, we haven't seen their exact building time, but we know that it should be definitely after $t_1$. For such samples, that we call \emph{censored} samples, we set $t=t_1-(t_0+\Phi)$ that is equal to the length of the observation window $\Omega$, and set $y=0$ to indicate that the recorded time is, in fact, a lower bound on the true relationship building time. These type of samples are also of interest because their features will give us some information about their time falling after $t_1$. As a result, each data sample is associated with a triple $(\mb{x},y,t)$ representing its feature vector, observation status, and the time it takes for the target relationship to be formed, respectively.

\begin{table}[t]
    \centering
    \caption{Similarity Meta-Paths in Different Networks}
    \label{table:meta}
    \footnotesize
    \begin{tabular} {c c l}
        \toprule
        Network & Meta-Path & Semantic Meaning \\
        \midrule
        \multirow{8}{*}{\rotatebox{90}{DBLP}} 
        &&\\
        & $A\rightarrow P\leftarrow A$ & Authors co-write a paper\\
        & $A\rightarrow P\leftarrow A\rightarrow P\leftarrow A$ & Authors have common co-author\\
        & $A\rightarrow P\leftarrow V\rightarrow P\leftarrow A$ & Authors publish in the same venue\\
        & $A\rightarrow P\rightarrow T\leftarrow P\leftarrow A$ & Authors use the same term\\
        & $A\rightarrow P\rightarrow P\leftarrow P\leftarrow A$ & Authors cite the same paper\\
        & $A\rightarrow P\leftarrow P\rightarrow P\leftarrow A$ & Authors are cited by the same paper\\
        &&\\
        \midrule
        \multirow{5}{*}{\rotatebox{90}{Delicious}} 
        &&\\
        & $U\leftrightarrow U\leftrightarrow U$ & Users have common contact\\
        & $U\rightarrow B\leftarrow U$ & Users post the same bookmark\\
        & $U\rightarrow B\rightarrow T\leftarrow B\leftarrow U$ & Users post bookmarks with the same tag\\
        &&\\
        \midrule
        \multirow{13}{*}{\rotatebox{90}{MovieLens}} 
        &&\\
        & $M\rightarrow A\leftarrow M$ & Movies share an actor\\
        & $M\rightarrow C\leftarrow M$ & Movies belong to the same country\\
        & $M\rightarrow D\leftarrow M$ & Movies have the same director\\
        & $M\rightarrow G\leftarrow M$ & Movies have the same genre\\
        & $M\rightarrow T\leftarrow M$ & Movies have the same tag\\
        & $U\rightarrow M\leftarrow U$ & Users rate common movie\\
        & $U\rightarrow M\rightarrow A\leftarrow M\leftarrow U$ & Users rate movies sharing an actor\\
        & $U\rightarrow M\rightarrow C\leftarrow M\leftarrow U$ & Users rate movies from the same country\\
        & $U\rightarrow M\rightarrow D\leftarrow M\leftarrow U$ & Users rate movies of the same director\\
        & $U\rightarrow M\rightarrow G\leftarrow M\leftarrow U$ & Users rate movies with the same genre\\
        & $U\rightarrow M\rightarrow T\leftarrow M\leftarrow U$ & Users rate movies with the same tag\\
        &&\\
        \bottomrule
    \end{tabular}
\end{table}

\subsection{Dynamic Feature Extraction}
In this part, we describe how to utilize the temporal history of the network in the feature extraction window in order to extract features for continuous-time relationship prediction problem. We first begin with the meta-path-based feature set for heterogeneous information networks, and then incorporate these features into a \emph{recurrent neural network based autoencoder} to exploit the temporal dynamics of the network as well. Hereby, we begin by defining the concept of meta-path \cite{sun2011pathsim}:

\begin{definition}[Meta-Path]
    In a heterogeneous information network, a meta-path is a directed path following the graph of the network schema to describe the general relations that can be derived from the network. Formally speaking, given a network schema $\mc{S}_G=(\mc{V}, \mc{E})$, the sequence $\nu_1\xrightarrow{\varepsilon_1}\nu_2\xrightarrow{\varepsilon_2}\dots\nu_{k-1}\xrightarrow{\varepsilon_{k-1}}\nu_k$ is a meta-path defined on $S_G$ where $\nu_i\in \mc{V}$ and $\varepsilon_i\in \mc{E}$.
\end{definition} 

Meta-paths are commonly used in heterogeneous information networks to describe multi-typed relations that have concrete semantic meanings. For example, in the bibliographic network whose schema is shown in Fig.~\ref{fig:schema:dblp}, we can define the co-authorship relation by the following meta-path:
\[Author\xrightarrow{write}Paper\xleftarrow{write}Author\]
or simply by $A\rightarrow P\leftarrow A$. Another example is the author citation relation, which in this paper is used as the target relation for DBLP network. It can be specified as:
\[Author\xrightarrow{write}Paper\xrightarrow{cite}Paper\xleftarrow{write}Author\]
abbreviated as $A\rightarrow P\rightarrow P\leftarrow A$.

We can extend the concept of the heterogeneous adjacency matrix, which is used to indicate relationships between nodes of different types, to \emph{meta-path adjacency matrix}, which we will use to indicate the number of path instances between two nodes of (possibly) different types, as explained below.

\begin{definition}{(Meta-path Adjacency Matrix)}
Given a heterogeneous network $G$ with schema $\mc{S}_G=(\mc{V}, \mc{E})$, and the meta-path $\nu_1\xrightarrow{\varepsilon_1}\nu_2\xrightarrow{\varepsilon_2}\dots\nu_{k-1}\xrightarrow{\varepsilon_{k-1}}\nu_k$ defined over $\mc{S}_G$ denoting the relation between node types $\nu_i,\nu_j\in\mc{V}$, the meta-path adjacency matrix $M_{\Psi}$ is defined as:
\[M_\Psi=\prod_{i=1}^{k-1}M_{\varepsilon_i}\]
which indicates the number of path instances between any node pair $u\in\nu_1$ and $v\in\nu_k$ following the meta-path $\Psi$. The time-aware counterpart of meta-path adjacency matrix is defined analogously by using the time-aware heterogeneous adjacency matrix.
\end{definition}

Among the possible meta-paths that can be defined on a network schema, there are some that capture the similarity between two nodes. For example, the co-authorship meta-path $A\rightarrow P\leftarrow A$ in a bibliographic network creates a sense of similarity between two \emph{Author} nodes. These type of meta-paths, called \emph{similarity meta-paths}, are widely used to define topological features for link prediction problem in heterogeneous networks \cite{sun2011co, zhang2014meta, 7752228}. Table~\ref{table:meta} presents a number of similarity meta-paths that can be defined on DBLP, Delicious, and MovieLens networks to capture the heterogeneous similarity between different node types.

The concept of similarity meta-paths can be extended to define heterogeneous features suitable for relationship prediction problem, where we have a target relation. Here we follow the same approach as in \cite{sun2012will} which suggests the following three meta-path-based building blocks to describe features for relationship prediction problem, given a target relation between two nodes of type $A$ and $B$:
\begin{enumerate}
    \small
    \item $A\xrsquigarrow{similarity}A\xrsquigarrow{target}B$
    \item $A\xrsquigarrow{target}B\xrsquigarrow{similarity}B$
    \item $A\xrsquigarrow{relation}C\xrsquigarrow{relation}B$
\end{enumerate}
where $\rightsquigarrow$ denotes a meta-path, with labels \emph{similarity} and \emph{target} denoting a similarity meta-path and the target relation, respectively. The \emph{relation} label denotes an arbitrary meta-path relating two nodes of possibly different types. The first block tells that there are some nodes of type $A$ similar to a single node of the same type that has made the target relationship with a node of type $B$. Therefore, those similar nodes may also form the target relation with the type $B$ node. An analogous intuition is behind the second block. For the third, it says that some nodes of type $A$ are in relation with some type $C$ nodes, which are themselves in relation with some nodes of type $B$. Hence, it is likely that type $A$ nodes form some relationships, such as the target relationship, with type $B$ nodes. We refer to the meta-paths that are created using these three blocks as \emph{feature meta-paths}.

As an example in DBLP bibliographic network, for the target relation, we use $A\rightarrow P\rightarrow P\leftarrow A$ as a meta-path denoting the author citation relation. In Addition, Paper-cite-Author ($P\rightarrow P\rightarrow A$) and Author-cite-Paper ($A\rightarrow P\rightarrow P$) are also used as the arbitrary relations, and the similarity meta-paths for DBLP network from Table~\ref{table:meta} are used to define the features for author citation relationship prediction.

After specifying feature meta-paths, we need a method to quantify them as numeric features. Due to the dynamicity of the network, different links are emerging and vanishing from the network over time. Therefore, the quantifying method must handle this dynamicity. Here, we formally define \emph{Time-Aware Meta-Path-based Features}:

\begin{definition}[Time-Aware Meta-Path-based Feature]
    Suppose that we are given a dynamic heterogeneous network $G^{\tau}$ along with its network schema $\mc{S}_G=(\mc{V}, \mc{E})$, and a target Relation $A\rightsquigarrow B$. For a given pair of nodes $a\in A$ and $b\in B$, and a feature meta-path $\Psi=A\xrightarrow{\varepsilon_1}\nu_1\xrightarrow{\varepsilon_2}\dots\nu_{n-1}\xrightarrow{\varepsilon_{n}}B$ defined on $\mc{S}_G$, the time-aware meta-path-based feature at the timestamp $\tau$ is the number of path instances between $a$ and $b$ following $\Psi$:
    \begin{equation*}
        f_{\Psi}^\tau(a,b)=M^\tau_{\Psi}[a,b]
    \end{equation*}
\end{definition}

This way, for any pair of nodes, we can quantify the number of path instances of a particular meta-path at any specific timestamp $\tau$. Although this quantification requires matrix multiplication, it can be done efficiently due to the following reasons:
\begin{enumerate}
\item The heterogeneous adjacency matrices are highly sparse, thus for calculating meta-path adjacency matrices, we can considerably reduce the time complexity of each single matrix multiplication by using fast sparse matrix multiplication algorithms \cite{horowitz1983fundamentals}.
\item The process of calculating the meta-path adjacency matrices is highly parallelizable, as the corresponding meta-paths decouples into simpler similarity meta-paths, which themselves decouple further into link types. Therefore, we can calculate the adjacency matrix of different similarity meta-paths in parallel, and then multiply them together to obtain the feature meta-path adjacency matrices.
\item Due to the similarity meta-paths sharing common sub-paths, computation time for the similarity meta-paths can also be saved using dynamic programming to avoid recalculating previously computed products. For example, for the DBLP dataset, if the target relation is $A\rightarrow P\rightarrow P\leftarrow A$, then by using the similarity meta-paths shown in the Table~\ref{table:meta}, the path $A\rightarrow P\rightarrow P$ will appear in all the following feature meta-paths:
\[A\rightarrow P\rightarrow P\leftarrow P\leftarrow A\]
\[A\rightarrow P\rightarrow P\rightarrow P\leftarrow A\]
\[A\rightarrow P\rightarrow P\leftarrow A\]
Therefore, we can calculate $M_{A\rightarrow P\rightarrow P}$ once and then reuse it in the calculation of the adjacency matrices of the above meta-paths.

\item Finally, the symmetry of the similarity meta-paths further reduces the number of products, because we can calculate the matrix corresponding to half of the path, and then multiply the resulting matrix by its transpose. For instance, the adjacency matrix of the similarity meta-path $A\rightarrow P\leftarrow V\rightarrow P\leftarrow A$ can be calculated as $X\cdot X^T$ where $X=M_{\text{write}}\cdot M_{\text{publish}}$, reducing the number of multiplications from three to two.
\end{enumerate}

So far we proposed a method to calculate the time-aware meta-path-based features, which is the number of path instances of a particular meta-path at the timestamp $\tau$. If we set this timestamp to the end of the feature extraction window, it is as though we are aggregating the whole network into a single snapshot observed at time $t_0+\Phi$. In order to avoid such an aggregation, we divide the feature extraction window into a sequence of $k$ contiguous intervals of a constant size $\Delta$, as shown in Fig.~\ref{fig:timeline}. By doing so, we intend to extract time-aware features in each sub-window that results in a multivariate time series containing the information about the temporal evolution of the topological features between any pair of nodes. With this in mind, we define \emph{Dynamic Meta-Path-based Time Series} as follows:

\begin{definition}[Dynamic Meta-Path-based Time Series]
    Suppose that we are given a dynamic heterogeneous network $G^{\tau}$ observed in a feature extraction window of size $\Phi$ ($t_0<\tau \le t_0+\Phi$), along with its network schema $\mc{S}_G=(\mc{V}, \mc{E})$ and a target relation $A\rightsquigarrow B$. Also suppose that the feature extraction window is divided into $k$ fragments of size $\Delta$. For a given pair of nodes $a\in A$ and $b\in B$ in $G^{t_0+\Phi}$, and a meta-path $\Psi$ defined on $\mc{S}_G$, the dynamic meta-path-based time series of $(a,b)$ is calculated as:
    \begin{equation*}
        x_{\Psi}^i(a,b)=f_{\Psi}^{t_0+i\Delta}(a,b) - f_{\Psi}^{t_0+(i-1)\Delta}(a,b)\quad\quad i=1\dots k
    \end{equation*}
\end{definition}

For each feature meta-path designed using the triple building blocks described before, we get a unique time series. For each time step, we put the corresponding values from all the time series into a vector. Consequently, we get a multivariate time series where each time step is vector-valued. For example, if we have $d$ feature meta-paths $\Psi_1$ to $\Psi_d$, then each time step of the resulting time series for any node pair $(a,b)$ will become:
\[\mb{x}^i_{a,b}=[x_{\Psi_1}^i(a,b),\dots,x_{\Psi_d}^i(a,b)]^T,\quad i=1\dots k\]
We refer to this vector-valued time series as \emph{Multivariate Meta-Path-based Time Series}. Such multivariate time series reflect how topological features change between two nodes across different snapshots of the network. Based on the level of the network dynamicity, it can capture increasing/decreasing trends or even periodic/re-occurring patterns.

Now it's time to convert the multivariate meta-path-based time series into a single feature vector so that we can use it as the input to our non-parametric model that will be discussed in the next section. A trivial solution would be to stack all the vector-valued time steps of the multivariate time series into a single vector. However, this approach will result in a very high dimensional vector as the number of time steps increases and can lead to difficulties in the learning procedure due to the curse of dimensionality. This is in contrast with our expectation that more time steps would bring more information about the history of the network and should result in a better prediction model. To overcome this problem, we combine the power of recurrent neural networks, especially Long Short Term Memory (LSTM) units \cite{hochreiter1997long}, which have proven to be very successful in handling time series and sequential data, with Autoencoders \cite{bengio2009learning}, which are widely used to learn alternative representations of the data such that the learned representation can reconstruct the original input. Our goal is to transform the multivariate meta-path-based time series into a compact vector representation such that the resulting vector holds as much information from the original multivariate time series as possible.

\begin{figure}
    \centering
    \footnotesize
    \tikzstyle{block} = [rectangle,draw=black,minimum width=0.5cm, minimum height=0.25cm]
    \tikzstyle{arrow} = [thick,->,>=stealth]
    \tikzstyle{label} = [rectangle]
    \begin{tikzpicture}
    \node[block] (e1) at (0,0) {};
    \node[block] (e2) at (1,0) {};
    \node[block,draw=none] (ed) at (2,0) {$\dots$};
    \node[block] (ek) at (3,0) {};
    
    \node[block] (dk) at (4,0) {};
    \node[block] (dk1) at (5,0) {};
    \node[block,draw=none] (dd) at (6,0) {$\dots$};
    \node[block] (d1) at (7,0) {};
    
    \node[label] (ie1) at (0,-1) {${x}^1$};
    \node[label] (ie2) at (1,-1) {${x}^2$};
    \node[label] (iek) at (3,-1) {${x}^k$};
    
    \node[label] (idk) at (4,-1) {$\mb{x}$};
    \node[label] (idk1) at (5,-1) {$\mb{x}$};
    \node[label] (id1) at (7,-1) {$\mb{x}$};
    
    \node[label] (oek) at (3,1) {$\mb{x}$};
    \node[label] (odk) at (4,1) {${x}^k$};
    \node[label] (odk1) at (5,1) {${x}^{k-1}$};
    \node[label] (od1) at (7,1) {${x}^1$};
    
    \draw [arrow] (ie1) -- (e1);
    \draw [arrow] (ie2) -- (e2);
    \draw [arrow] (iek) -- (ek);
    
    \draw [arrow] (idk) -- (dk);
    \draw [arrow] (idk1) -- (dk1);
    \draw [arrow] (id1) -- (d1);
    
    \draw [arrow] (ek) -- (oek);
    \draw [arrow] (dk) -- (odk);
    \draw [arrow] (dk1) -- (odk1);
    \draw [arrow] (d1) -- (od1);
    
    \draw [arrow] (e1) -- (e2);
    \draw [arrow] (e2) -- (ed);
    \draw [arrow] (ed) -- (ek);
    \draw [arrow] (ek) -- (dk);
    \draw [arrow] (dk) -- (dk1);
    \draw [arrow] (dk1) -- (dd);
    \draw [arrow] (dd) -- (d1);
    
    \end{tikzpicture}
    \caption{The architecture of the LSTM Autoencoder used for dynamic feature extraction. The first $k$ steps depicts the manner of the working of the encoder LSTM, while the second $k$ steps describes the decoder LSTM. The output of the $k^{\text{th}}$ stage is used as the feature vector $\mb{x}$, which is fed into the decoder $k$ times to produce the input sequence in the reversed order.}
    \label{fig:autoencoder}
\end{figure}
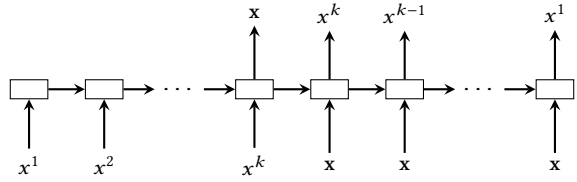

Inspired by the work of Dai and Le on semi-supervised sequence learning \cite{dai2015semi}, we design an autoencoder that learns how to take a multivariate time series as input and compress it into a latent vector representation. The architecture of such autoencoder is illustrated in Fig.~\ref{fig:autoencoder}. The autoencoder consists of two components: (1) the encoder, which takes the input data and transforms it into a latent representation; and (2) the decoder, which takes the encoded representation and transforms it back to the input space. The autoencoder is trained in such a way that it can reconstruct the original input data. 

As the purpose of using the autoencoder in this paper is to compress multivariate time series, instead of using simple feed-forward neural networks, both encoder and decoder are built using LSTMs. The input to the encoder LSTM is a multivariate time series of length $k$. The encoder accepts the vector-valued time steps of the input multivariate time series sequentially. After receiving the $k^{\text{th}}$ time step, the output of the encoder LSTM will be the compressed feature vector that we will use as the input to the \npglm method. In order to train the encoder to learn how to compress the input time series, it is matched with a decoder LSTM. The decoder LSTM receives $k$ copies of the compressed feature vector one after another, and with a proper loss function (such as mean squared error) it is forced to reconstruct the original multivariate time series in reverse order. Reversing the output sequence will make the optimization of the model easier since it causes the decoder to revert back the changes made by the encoder to the input sequence.

The benefits of using the LSTM autoencoder is three-fold: (1) since the autoencoder can reconstruct the original time series, which reflects the temporal dynamics of the network, we get minimum information loss in the compressed feature vector; (2) as we can set the dimensionality of the compressed feature vector to any desired value, we can evade the curse of dimensionality; and (3) due to the inherent dynamicity of recurrent neural networks and LSTMs, when we receive $(k+1)^{\text{th}}$ snapshot of the network, we can easily fine-tune the previous autoencoder that was learned with $k$ snapshots to consider the new snapshot as well, instead of repeating the whole learning procedure from scratch.

To conclude this section, we quickly review the whole procedure of processing the network data, training the autoencoder, and assembling a training dataset for the supervised model to predict the building time of a particular target relation:
\begin{enumerate}
    \item The network evolution timeline is split into the feature extraction window and the observation window.
    \item Those node pairs that have either formed the target relationship in the observation window (observed samples) or have not formed the target relationship at all (censored samples) are selected as sample node pairs.
    \item By extracting feature meta-paths based on the target relation and similarity meta-paths, a multivariate time series can be obtained for each sample node pair. Thus if we have $N$ sample node pairs, we will have a dataset of $N$ multivariate time series.
    \item The LSTM autoencoder is trained using the dataset of $N$ multivariate time series to learn how to compress time series into feature vectors.
    \item For each sample node pair, the corresponding multivariate time series is compressed into a feature vector $\mb{x}$ using the learned encoder LSTM. 
    \item For each observed node pair, the feature vector $\mb{x}$ is labeled with $y=1$ and associates with the variable $t$ denoting the time it takes for the node pair to form the target relationship. For censored node pairs, $y$ is set to zero and $t$ becomes equal to the size of the observation window.
    \item Finally, we will have a dataset of the form $\{\mb{x},y,t\}_i,\ i=1\dots N$ that will be used to train the supervised model.
\end{enumerate}

We explain our proposed non-parametric model in the next section that takes the learned representation as the feature vector $\mb{x}$ and attempts to predict the corresponding event time $t$.

\begin{table*}
    \centering
    \caption{Characteristics of Some Probability Distributions Used for Event-Time Modeling}
    \label{table:dists}
    \footnotesize
    \begin{tabu} to \textwidth {X X[c] X[c] X[c] X[c]}
        \toprule
        Distribution & Density function & Survival function & Intensity function & Cumulative intensity\\
        & $f_T(t)$ & $S(t)$ & $\lambda(t)$ & $\Lambda(t)$\\[1pt]
        \midrule 
        Exponential & $\alpha\exp(-\alpha t)$ & $\exp(-\alpha t)$ & $\alpha$ & $\alpha t$\\[4pt]
        Rayleigh & $\frac{t}{\sigma^2}\exp(-\frac{t^2}{2\sigma^2})$ & $\exp(-\frac{t^2}{2\sigma^2})$ & $\frac{t}{\sigma^2}$ & $\frac{t^2}{2\sigma^2}$\\[4pt]
        Gompertz & $\alpha e^t\exp\left\lbrace -\alpha(e^t-1) \right\rbrace$ & $\exp\left\lbrace -\alpha(e^t-1) \right\rbrace$ & $\alpha e^t$ & $\alpha e^t$\\[4pt]
        Weibull & $\frac{\alpha t^{\alpha-1}}{\beta^\alpha}\exp\left\lbrace-(\frac{t}{\beta})^\alpha\right\rbrace$ & $\exp\left\lbrace-(\frac{t}{\beta})^\alpha\right\rbrace$ & $\frac{\alpha t^{\alpha-1}}{\beta^\alpha}$ & $(\frac{t}{\beta})^\alpha$\\[2pt]
        \bottomrule 
    \end{tabu}
\end{table*}

\section{Supervised Non-Parametric Model}\label{sec:method}
In this section we introduce our proposed model, called \emph{Non-Parametric Generalized Linear Model}, to solve the problem of continuous-time relationship prediction based on the extracted features. 
Since the relationship building time is treated as a continuous random variable, we attempt to model the probability distribution of this time, given the features of the target relationship. Thus, if we denote the target relationship building time by $t$ and its features by $\mb{x}$, our aim is to model the probability density function $f_T(t\mid \mb{x})$. A conventional approach to modeling this function is to fix a parametric distribution for $t$ (e.g. Exponential distribution) and then relate $\mb{x}$ to $t$ using a Generalized Linear Model \cite{sun2012will}. The major drawback of this approach is that we need to know the exact distribution of the relationship building time, or at least, we could guess the best one that fits. The alternative way that we follow is to learn the shape of $f_T(t\mid \mb{x})$ from the data using a non-parametric solution.

In the rest of this section, we first bring the necessary theoretical backgrounds related to the concept, then we go through the details of the proposed model. In the end, we explain the learning and inference algorithms of \npglm.

\subsection{Background}
Here we define some essential concepts that are necessary to study before we proceed to the proposed model. Generally, the formation of a relationship between two nodes in a network can simply be considered as an event with its occurring time as a random variable $T$ coming from a density function $f_T(t)$. Regarding this, we can have the following definitions:

\begin{definition}[Survival Function]
    Given the density $f_T(t)$, the survival function denoted by $S(t)$, is the probability that an event occurs after a certain value of $t$, which means:
    \begin{equation}
    S(t) = P(T > t) = \int_t^\infty f_T(t)dt
    \end{equation}
\end{definition}

\begin{definition}[Intensity Function]
    The intensity function (or failure rate function), denoted by $\lambda(t)$, is the instantaneous rate of event occurring at any time $t$ given the fact that the event has not occurred yet:
    \begin{equation}
    \lambda(t)=\lim_{\Delta t\rightarrow 0}\frac{P(t\le T\le t+\Delta t\mid T\ge t)}{\Delta t}
    \end{equation}
\end{definition}

\begin{definition}[Cumulative Intensity Function]
    The cumulative intensity function, denoted by $\Lambda(t)$, is the area under the intensity function up to a point $t$:
    \begin{equation}\label{eq:Lambda}
    \Lambda(t)=\int_0^t\lambda(t)dt
    \end{equation}
\end{definition}

The relations between density, survival, and intensity functions come directly from their definitions as follows:

\begin{equation}\label{eq:intensity}
\lambda(t)=\frac{f_T(t)}{S(t)}
\end{equation}
\begin{equation}\label{eq:reliability}
S(t)=\exp(-\Lambda(t))
\end{equation}
\begin{equation}\label{eq:density}
f_T(t)=\lambda(t)\exp(-\Lambda(t))
\end{equation}

Table \ref{table:dists} shows the density, survival, intensity, and cumulative intensity functions of some widely-used distributions for event time modeling.

\subsection{Model Description}
Looking at Eq.~\ref{eq:density}, we see that the density function can be specified uniquely with its intensity function. Since the intensity function often has a simpler form than the density itself, if we learn the shape of the intensity function, then we can infer the entire distribution eventually. Therefore, we focus on learning the shape of the conditional intensity function $\lambda(t\mid \mb{x})$ from the data, and then accordingly infer the conditional density function $f_T(t\mid \mb{x})$ based on the learned intensity.
In order to reduce the hypothesis space of the problem and avoid the curse of dimensionality, we assume that $\lambda(t\mid \mb{x})$, which is a function of both $t$ and $\mb{x}$, can be factorized into two separate positive functions as the following:
\begin{equation}\label{eq:lambda}
\lambda(t\mid \mb{x})=g(\mb{w}^T\mb{x})h(t)
\end{equation}
where $g$ is a function of $\mb{x}$ which captures the effect of features via a linear transformation using coefficient vector $\mb{w}$ independent of $t$, and $h$ is a function of $t$ which captures the effect of time independent of $x$. This assumption, referred to as \emph{proportional hazards condition} \cite{breslow1975analysis}, holds in GLM formulations of many event-time modeling distributions, such as the ones shown in Table~\ref{table:dists}. Our goal is now to fix the function $g$ and then learn both the coefficient vector $\mb{w}$ and the function $h$ from the training data. In order to do so, we begin with the likelihood function of the data which can be written as follows:

\begin{equation}
\prod_{i=1}^{N}f_T(t_i\mid \mb{x}_i)^{y_i}P(T\ge t_i\mid \mb{x}_i)^{1-y_i}\\
\end{equation}
The likelihood consists of the product of two parts: The first part is the contribution of those samples for which we have observed their exact building time, in terms of their density function. The second part on the other hand, is the contribution of the censored samples, for which we use the probability of the building time being greater than the recorded one. By applying Eq.~\ref{eq:reliability} and \ref{eq:density} we can write the likelihood in terms of the intensity function:
\begin{equation}
\prod_{i=1}^{N}\big[\lambda(t_i\mid x_i)\exp\{-\Lambda(t_i\mid x_i)\}\big]^{y_i}\exp\lbrace-\Lambda(t_i\mid x_i)\rbrace^{1-y_i}
\end{equation}
By merging the exponentials and applying Eq.~\ref{eq:Lambda} and \ref{eq:lambda}, the likelihood function becomes:
\begin{equation}
\prod_{i=1}^{N}\left[g(\mb{w}^T\mb{x}_i)h(t_i)\right]^{y_i}\exp\lbrace-g(\mb{w}^T\mb{x}_i)\int_{0}^{t_i}h(t)dt\rbrace
\end{equation}

Since we don't know the form of $h(t)$, we cannot directly calculate the integral appeared in the likelihood function. To deal with this problem, we treat $h(t)$ as a non-parametric function by approximating it with a piecewise constant function that changes just in $t_i$s. Therefore, the integral over $h(t)$, denoted by $H(t)$, becomes a series:
\begin{equation}\label{eq:cumh}
H(t_i)=\int_{0}^{t_i}h(t)dt \simeq \sum_{j=1}^{i}h(t_j)(t_j-t_{j-1})
\end{equation}
assuming samples are sorted by $t$ in increasing order, without loss of generality. The function $H(t)$ defined above plays an important role in both learning and inference phases. In fact, both the learning and inference phases rely on $H(t)$ instead of $h(t)$, which we will see later in this paper.
Replacing the above series in the likelihood, taking the logarithm and negating, we end up with the following negative log-likelihood function, simply called the \emph{loss function}, denoted by $L$:

\begin{equation}\label{eq:logl}
\begin{split}
L(\mb{w},h)
=\sum_{i=1}^{N}\Big\lbrace g(\mb{w}^T\mb{x}_i)\sum_{j=1}^{i}h(t_j)(t_j-t_{j-1})-y_i\left[\log g(\mb{w}^T\mb{x}_i) + \log h(t_i)\right]\Big\rbrace\\
\end{split}
\end{equation}

The loss function depends on both the vector $\mb{w}$ and the function $h(t)$. In the next part, we explain an iterative learning algorithm to learn both $\mb{w}$ and $h(t)$ collectively.

\subsection{Learning Algorithm}
Minimizing the loss function (Eq.~\ref{eq:logl}) relies on the choice of the function $g$. There are no particular limits on the choice of $g$ except that it must be a non-negative function. For example, both quadratic and exponential functions of $\mb{w}^T\mb{x}$ will do the trick. Here, we proceed with $g(\mb{w}^T\mb{x})=\exp(\mb{w}^T\mb{x})$ since it makes the loss function convex with respect to $\mb{w}$. Subsequent equations can be derived for other choices of $g$ analogously.

Setting the derivative of the loss function with respect to $h(t_k)$ to zero yields a closed form solution for $h(t_k)$:
\begin{equation}\label{eq:h}
h(t_k)=\frac{y_k}{(t_k-t_{k-1})\sum_{i=k}^{N}\exp(\mb{w}^T\mb{x}_i)}
\end{equation}

By applying Eq.~\ref{eq:cumh}, we get the following for $H(t_i)$:
\begin{equation}\label{eq:H}
H(t_i)=\sum_{j=1}^{i}\frac{y_j}{\sum_{k=j}^{N}\exp(\mb{w}^T\mb{x}_k)}
\end{equation}
which depends on the vector $\mb{w}$. On the other hand, we cannot obtain a closed form solution for $\mb{w}$ from the loss function. Therefore, we turn to use Gradient-based optimization methods to find the optimal value of $\mb{w}$. The loss function with respect to $\mb{w}$ is as follows:

\begin{equation}\label{eq:nlw}
L(\mb{w})=\sum_{i=1}^{N}\left\lbrace\exp(\mb{w}^T\mb{x}_i)H(t_i)-y_i\mb{w}^T\mb{x}_i\right\rbrace + Const.
\end{equation}
which depends on the function $H$. As the learning of both $\mb{w}$ and $H$ depends on each other, they should be learned collectively. Here, we use an iterative algorithm to learn $\mb{w}$ and $H$ alternatively. We begin with a random vector $\mb{w}^{(0)}$. Then in each iteration $\tau$, we first update $H^{(\tau)}$ via Eq.~\ref{eq:H} using $w^{(\tau-1)}$. Next, we optimize Eq.~\ref{eq:nlw} using the values of $H^{(\tau)}(t_i)$ to obtain $\mb{w}^{(\tau)}$. We continue this routine until convergence. Since this procedure successively reduces the value of the loss function, and as the loss function (i.e. the negative log-likelihood) is bounded from below, the algorithm will ultimately converge to a stationary point. The pseudo code of the learning procedure is given in Algorithm~\ref{alg:learning}.

\begin{algorithm}[t]
    \small
    \SetAlgoLined
    \KwIn{$\mb{X}_{N\times d}=(\mb{x}_1,\dots\mb{x}_N)^T$ as $d$-dimensional feature vectors, $\mb{y}_{N\times1}$ as observation states, and $\mb{t}_{N\times1}$ as recorded times.}
    \KwOut{Learned parameters $\mb{w}_{d\times1}$ and $\mb{H}_{N\times1}$.}
    $converged\leftarrow False$\;
    $threshold\leftarrow10^{-4}$\;
    $\tau\leftarrow 0$\;
    $L^{(\tau)}=\infty$\;
    Initialize $\mb{w}^{(\tau)}$ with random values\;
    \While{Not $converged$}{
        $\tau\leftarrow\tau+1$\;
        Use Eq.~\ref{eq:H} to obtain $\mb{H}^{(\tau)}$ using $\mb{w}^{(\tau-1)}$\;
        Minimize Eq.~\ref{eq:nlw} to obtain $\mb{w}^{(\tau)}$ using $\mb{H}^{(\tau)}$\;
        Use Eq.~\ref{eq:logl} to obtain $L^{(\tau)}$ using $\mb{w}^{(\tau)}$ and $\mb{H}^{(\tau)}$\;
        
        \If{$\left\|{L}^{(\tau)} - {L}^{(\tau-1)}\right\| < threshold$}{
            $converged\leftarrow True$\;
        }
    }
    $\mb{w}\leftarrow \mb{w}^{(\tau)}$\;
    $\mb{H}\leftarrow \mb{H}^{(\tau)}$\;
    \caption{The learning algorithm of \npglm}
    \label{alg:learning}
\end{algorithm}

\subsection{Inference Queries}
In this part, we explain how to answer the common inference queries based on the inferred distribution $f_T(t\mid \mb{x})$. Suppose that we have learned the vector ${\mb{w}}$ and the function ${H}$ using the training samples $(\mb{x}_i, y_i, t_i),\ i=1\dots N$ following Algorithm~\ref{alg:learning}. Afterward, for a testing relationship $R$ associated with a feature vector $\mb{x}_R$, the following queries can be answered:\\

\subsubsection{Ranged Probability} What is the probability for the relationship $R$ to be formed between time $t_\alpha$ and $t_\beta$? This is equivalent to calculating $P(t_\alpha \le T \le t_\beta \mid \mb{x}_R)$, which by definition is:
\begin{equation}\label{eq:ranged}
\begin{split}
P(t_\alpha\le T \le t_\beta \mid \mb{x}_R) = S(t_\alpha\mid \mb{x}_R) - S(t_\beta\mid \mb{x}_R)\\
= \exp\{-g(\mb{w}^T\mb{x}_R){H}(t_\alpha)\} - \exp\{-g(\mb{w}^T\mb{x}_R){H}(t_\beta)\}
\end{split}
\end{equation}
The problem here is to obtain the values of ${H}(t_\alpha)$ and ${H}(t_\beta)$, as $t_\alpha$ and $t_\beta$ may not be among $t_i$s of the training samples, for which ${H}$ is estimated. To calculate ${H}(t_\alpha)$, we find $k\in\{1,2,\dots,N\}$ such that $t_k\le t_\alpha < t_{k+1}$.
Due to the piecewise constant assumption for the function $h$, we get:
\begin{equation}\label{eq:inf1}
{h}(t_\alpha)=\frac{{H}(t_\alpha)-{H}(t_k)}{t_\alpha-t_k}
\end{equation} 
On the other hand, since $h$ only changes in $t_i$s, we have:
\begin{equation}\label{eq:inf2}
{h}(t_\alpha)={h}(t_{k+1})=\frac{{H}(t_{k+1})-{H}(t_k)}{t_{k+1}-t_k}
\end{equation}
Combining Eq.~\ref{eq:inf1} and \ref{eq:inf2}, we get:
\begin{equation}\label{eq:inf3}
{H}(t_\alpha)={H}(t_k)+(t_\alpha-t_k)\frac{{H}(t_{k+1})-{H}(t_k)}{t_{k+1}-t_k}
\end{equation}
Following the similar approach, we can calculate ${H}(t_\beta)$, and then answer the query using Eq.~\ref{eq:ranged}. The dominating operation here is to find the value of $k$. Since we have $t_i$s sorted beforehand, this operation can be done using a binary search with $O(\log N)$ time complexity.\\

\subsubsection{Quantile} By how long the target relationship $R$ will be formed with probability $\alpha$? This question is equivalent to find the time $t_\alpha$ such that $P(T \le t_\alpha\mid x_R)=\alpha$. By definition, we have:
\begin{equation*}
\begin{split}
1-P(T \le t_\alpha\mid \mb{x}_R)=S(t_\alpha\mid \mb{x}_R)&=\exp\{-g(\mb{w}^T\mb{x}_R){H}(t_\alpha)\}=1-\alpha
\end{split}
\end{equation*}
Taking logarithm of both sides and rearranging, we get:
\begin{equation}\label{eq:inf4}
{H}(t_\alpha)=-\frac{\log(1-\alpha)}{g(\mb{w}^T\mb{x}_R)}
\end{equation}
To find $t_\alpha$, we first find $k$ such that ${H}(t_k)\le{H}(t_\alpha)<{H}(t_{k+1})$.
We eventually have $t_k\le t_\alpha < t_{k+1}$ since $H$ is a non-decreasing function due to non-negativity of the function $h$. Therefore, we again end up with Eq.~\ref{eq:inf3}, by rearranging which we get:
\begin{equation}\label{eq:inf5}
t_\alpha=(t_{k+1}-t_k)\frac{{H}(t_\alpha)-{H}(t_k)}{{H}(t_{k+1})-{H}(t_k)}+t_k
\end{equation}
By combining the Eq.~\ref{eq:inf4} and \ref{eq:inf5}, we can obtain the value of $t_\alpha$, which is the answer to the quantile query. It worth mentioning that if $\alpha=0.5$ then $t_\alpha$ becomes the median of the distribution $f_T(t\mid \mb{x}_R)$. Here again the dominant operation is to find the value of $k$, which due to the non-decreasing property of the function ${H}$ can be found using a binary search with $O(\log N)$ time complexity.

\subsubsection{Random Sampling}
Generating random samples from the inferred distribution can easily be carried out using the Inverse-Transform sampling algorithm. To pick a random sample from the inferred distribution $f_T(t\mid x)$, we first generate uniform random variable $u\sim Uniform(0,1)$. Then, we find $k$ such that $S(t_{k+1}\mid x)\leq u\le S(t_k\mid x)$. We output $t_{k+1}$ as the generated sample. Again, searching for the suitable value of $k$ is the dominant operation which can be undertaken via binary search with $O(\log N)$ time complexity.

\section{Synthetic Evaluations}\label{sec:synthetic}

\begin{algorithm}[t]
	\small
	\SetAlgoLined
	\KwIn{The number of observed samples $N_o$, the number of censored samples $N_c$, the dimension of the feature vectors $d$, and the desired distribution $dist$}
	\KwOut{Synthetically generated data $\mb{X}_{N\times d}$, $\mb{y}_{N\times1}$, and $\mb{t}_{N\times1}$.}
	$N\leftarrow N_o+N_c$\;
	Draw a weight vector $\mb{w}\sim\mathcal{N}(0,\mb{I}_d)$, where $\mb{I}_d$ is the $d$-dimensional identity matrix\;
	Draw scalar intercept $b\sim\mathcal{N}(0,1)$\;
	\For{$i\leftarrow1$ to $N$}{
		Draw feature vector $\mb{x}_i\sim\mathcal{N}(0,\mb{I}_d)$\;
		Set distribution parameter $\alpha_i\leftarrow\exp(\mb{w}^T\mb{x}_i+b)$\;
		\uIf{$dist == Rayleigh$}{
			Draw $t_i\sim\alpha_i~t\exp\{-0.5\alpha_it^2\}$\;
		}
		\uElseIf{$dist == Gompertz$}{
			Draw $t_i\sim\alpha_i~e^t\exp\{-\alpha_i(e^t-1)\}$\;
		}
	}
	
	Sort pairs $(\mb{x}_i,t_i)$ by $t_i$ in ascending order\;
	
	\For{$i\leftarrow1$ to $N_o$}{
		$y_i\leftarrow1$\;
	}
	\For{$i\leftarrow(N_o+1)$ to $N$}{
		$y_i\leftarrow0$\;
	}
	\caption{Synthetic dataset generation algorithm.}
	\label{alg:syn}
\end{algorithm}

We use synthetic data to verify the correctness of \npglm and its learning algorithm. Since \npglm is a non-parametric method, we generate synthetic data using various parametric models with previously known random parameters and evaluate how well \npglm can learn the parameters and the underlying distribution of the generated data.

\subsection{Experiment Setup}
We consider generalized linear models of two widely used distributions for event-time modeling, Rayleigh and Gompertz, as the ground truth models for generating synthetic data. Algorithm~\ref{alg:syn} is used to generate a total of $N$ data samples with $d$-dimensional feature vectors, consisting $N_o$ non-censored (observed) samples and remaining $N_c=N-N_o$ censored ones. For all synthetic experiments, we generate 10-dimensional feature vectors ($d=10$). We repeat every experiment 100 times and report the average results.

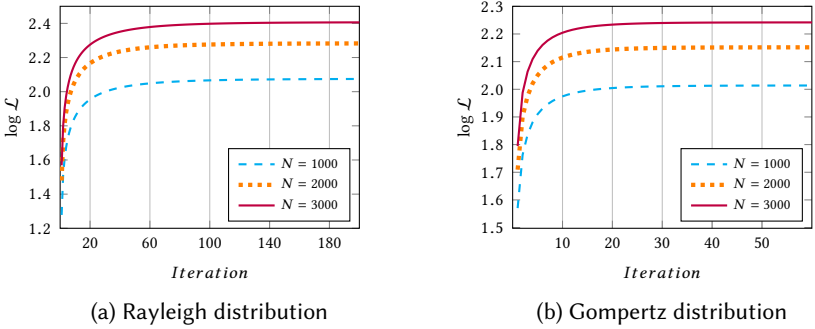
\begin{figure}[t]
	\subfloat[Rayleigh distribution]{
		\begin{tikzpicture}[trim axis left, trim axis right]
		\begin{axis}
		[
		tiny,
		width=0.4\columnwidth,
		height=4.5cm,
		legend pos=south east,
		legend style={font=\scriptsize,nodes={scale=0.75, transform shape}},
		xmajorgrids,
		y tick label style={
			/pgf/number format/.cd,
			fixed,
			fixed zerofill,
			precision=1,
			/tikz/.cd
		},
		xlabel=$Iteration$,
		ylabel=$\log\mathcal{L}$,
		ylabel shift = -4 pt,
		ymax=2.5,
		ymin=1.2,
		xmin=0,
		xmax=200,
		xtick={20,60,...,180},
		restrict x to domain=0:200,
		legend entries={${\tiny N=1000}$, $N=2000$, $N=3000$},
		]
		\addplot[color=cyan,  thick, dashed] table{results/cvg_ray_1000.txt};
		\addplot[color=orange,ultra thick, dotted] table{results/cvg_ray_2000.txt};
		\addplot[color=purple,thick] table{results/cvg_ray_3000.txt};
		\end{axis}
		\end{tikzpicture}
	}\hfil
	\subfloat[Gompertz distribution]{
		\begin{tikzpicture}[trim axis left, trim axis right]
		\begin{axis}
		[
		tiny,
		width=0.4\columnwidth,
		height=4.5cm,
		legend pos=south east,
		legend style={font=\scriptsize,nodes={scale=0.75, transform shape}},
		xmajorgrids,
		y tick label style={
			/pgf/number format/.cd,
			fixed,
			fixed zerofill,
			precision=1,
			/tikz/.cd
		},
		xlabel=$Iteration$,
		ylabel=$\log\mathcal{L}$,
		ylabel shift = -4 pt,
		ymax=2.3,
		xmin=0,
		xmax=60,
		xtick={10,20,...,50},
		restrict x to domain=0:100,
		legend entries={$N=1000$, $N=2000$, $N=3000$},
		]
		\addplot[color=cyan  ,thick, dashed] table{results/cvg_gom_1000.txt};
		\addplot[color=orange,ultra thick, dotted] table{results/cvg_gom_2000.txt};
		\addplot[color=purple,thick] table{results/cvg_gom_3000.txt};
		\end{axis}
		\end{tikzpicture}
	}
	\caption{Convergence of \npglm's average log-likelihood ($\log\mathcal{L}$) for different number of training samples ($N$). Censoring ratio has been set to 0.5.}
	\label{fig:syn-cvg-n}
\end{figure}

\begin{figure}[t]
	\subfloat[Rayleigh distribution]{
		\begin{tikzpicture}[trim axis left, trim axis right]
		\begin{axis}
		[
		tiny,
		width=0.4\columnwidth,
		height=4.5cm,
		legend pos=south east,
		legend style={font=\scriptsize,nodes={scale=0.75, transform shape}},
		xmajorgrids,
		y tick label style={
			/pgf/number format/.cd,
			fixed,
			fixed zerofill,
			precision=1,
			/tikz/.cd
		},
		xlabel=$Iteration$,
		ylabel=$\log\mathcal{L}$,
		ylabel shift = -8 pt,
		ymin=-2,
		xmin=0,
		xmax=100,
		xtick={10,30,...,90},
		restrict x to domain=0:200,
		legend entries={5\% censoring, 25\% censoring, 50\% censoring},
		]
		\addplot[color=cyan  ,thick, dashed] table{results/cvg_ray_5.txt};
		\addplot[color=orange,ultra thick, dotted] table{results/cvg_ray_25.txt};
		\addplot[color=purple,thick] table{results/cvg_ray_50.txt};
		\end{axis}
		\end{tikzpicture}
	}\hfil
	\subfloat[Gompertz distribution]{
		\begin{tikzpicture}[trim axis left, trim axis right]
		\begin{axis}
		[
		tiny,
		width=0.4\columnwidth,
		height=4.5cm,
		legend pos=south east,
		legend style={font=\scriptsize,nodes={scale=0.75, transform shape}},
		xmajorgrids,
		y tick label style={
			/pgf/number format/.cd,
			fixed,
			fixed zerofill,
			precision=1,
			/tikz/.cd
		},
		xlabel=$Iteration$,
		ylabel=$\log\mathcal{L}$,
		ylabel shift = -4 pt,
		xmin=0,
		xmax=60,
		xtick={10,20,...,50},
		restrict x to domain=0:60,
		legend entries={5\% censoring, 25\% censoring, 50\% censoring},
		]
		\addplot[color=cyan  ,thick, dashed] table{results/cvg_gom_5.txt};
		\addplot[color=orange,ultra thick, dotted] table{results/cvg_gom_25.txt};
		\addplot[color=purple,thick] table{results/cvg_gom_50.txt};
		\end{axis}
		\end{tikzpicture}
	}
	\caption{Convergence of \npglm's average log-likelihood ($\log\mathcal{L}$) for different censoring ratios with 1K samples.}
	\label{fig:syn-cvg-c}
\end{figure}
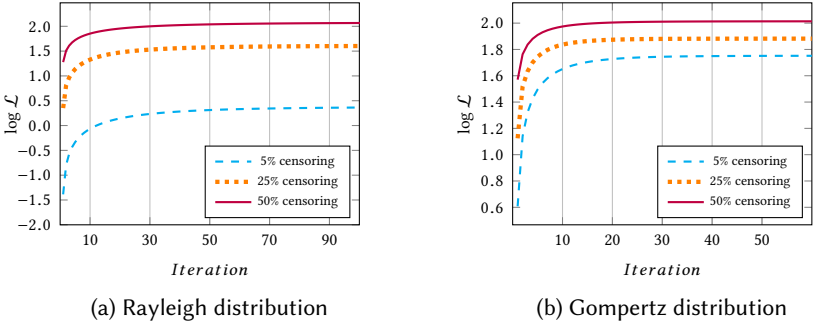

\subsection{Experiment Results}
\subsubsection{Convergence Analysis}
Since \npglm's learning is done in an iterative manner, we first analyze whether this algorithm converges as the number of iterations increases. We recorded the log-likelihood of \npglm, averaged over the number of training samples $N$ in each iteration. We repeated this experiment for $N\in\{1000,2000,3000\}$ with a fixed censoring ratio of 0.5, which means half of the samples are censored. The result is depicted in Fig.~\ref{fig:syn-cvg-n}. We can see that the algorithm successfully converges with a rate depending on the underlying distribution. For the case of Rayleigh, it requires about 100 iterations to converge but for Gompertz, this reduces to about 30. Also, we see that using more training data leads to achieving more log-likelihood as expected.

In Fig.~\ref{fig:syn-cvg-c}, we fixed $N=1000$ and performed the same experiment this time using different censoring ratios. According to the figure, we see that by increasing the censoring ratio, the convergence rate increases. This is because \npglm infers the values of $H(t)$ for all $t$ in the observation window. Therefore, as the censoring ratio increases, the observation window is decreased, so \npglm has to infer a fewer number of parameters, leading to a faster convergence. Note that as opposed to Fig.~\ref{fig:syn-cvg-n}, here a higher log-likelihood doesn't necessarily indicate a better fit, due to the likelihood marginalization we get by censored samples.

\subsubsection{Performance Analysis}
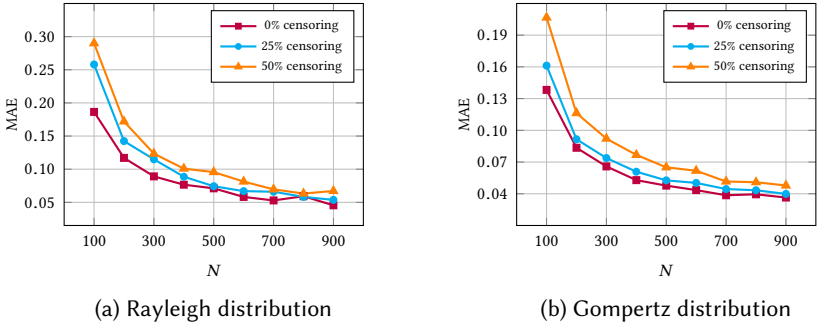
\begin{figure}[t]
	\subfloat[Rayleigh distribution]{
		\begin{tikzpicture}[trim axis left, trim axis right]
		\begin{axis}
		[
		tiny,
		width=0.4\columnwidth,
		height=4.5cm,
		legend pos=north east,
		legend style={font=\scriptsize,nodes={scale=0.75, transform shape}},
		grid,
		y tick label style={
			/pgf/number format/.cd,
			fixed,
			fixed zerofill,
			precision=2,
			/tikz/.cd
		},
		xlabel=$ N $,
		ylabel=MAE,
		ylabel shift = -4 pt,
		ymax=0.35,
		xmin=0,
		xmax=1000,
		ytick={0.05,0.10,...,0.35},
		xtick={100,300,...,900},
		restrict x to domain=0:900,
		legend entries={0\% censoring, 25\% censoring, 50\% censoring},
		]
		\addplot[color=purple,mark=square*,mark size=1.1,thick] table{results/mae_ray.txt};
		\addplot[color=cyan,mark=*,mark size=1.1,thick] table{results/mae_ray_25.txt};
		\addplot[color=orange,mark=triangle*,mark size=1.5,thick] table{results/mae_ray_50.txt};
		\end{axis}
		\end{tikzpicture}
	}\hfil
	\subfloat[Gompertz distribution]{
		\begin{tikzpicture}[trim axis left, trim axis right]
		\begin{axis}
		[
		tiny,
		width=0.4\columnwidth,
		height=4.5cm,
		legend pos=north east,
		legend style={font=\scriptsize,nodes={scale=0.75, transform shape}},
		grid,
		y tick label style={
			/pgf/number format/.cd,
			fixed,
			fixed zerofill,
			precision=2,
			/tikz/.cd
		},
		xlabel=$ N $,
		ylabel=MAE,
		ylabel shift = -4 pt,
		ymax=0.22,
		ymin=0.01,
		xmin=0,
		xmax=1000,
		xtick={100,300,...,900},
		restrict x to domain=0:900,
		ytick={0.04,0.07,...,0.21},
		legend entries={0\% censoring, 25\% censoring, 50\% censoring},
		]
		\addplot[color=purple,mark=square*,mark size=1.1,thick] table{results/mae_gom.txt};
		\addplot[color=cyan,mark=*,mark size=1.1,thick] table{results/mae_gom_25.txt};
		\addplot[color=orange,mark=triangle*,mark size=1.5,thick] table{results/mae_gom_50.txt};
		\end{axis}
		\end{tikzpicture}
	}
	\caption{\npglm's mean absolute error (MAE) vs the number of training samples ($N$) for different censoring ratios.}
	\label{fig:syn-mae-n}
\end{figure}

Next, we evaluated how good \npglm can infer the parameters used to generate synthetic data. To this end, we varied the number of training samples $N$ and measured the mean absolute error (MAE) between the learned weight vector $\hat{\mathbf{w}}$ and the ground truth. Fig.~\ref{fig:syn-mae-n} illustrates the result for different censoring ratios. It can be seen that as the number of training samples increases, the MAE gradually decreases. The other point to notice is that more censoring ratio results in a higher error due to the information loss we get by censoring.

\begin{figure}[t]
	\subfloat[Rayleigh distribution]{
		\begin{tikzpicture}[trim axis left, trim axis right]
		\begin{axis}
		[
		tiny,
		width=0.4\columnwidth,
		height=4.5cm,
		legend pos=north east,
		legend style={font=\scriptsize,nodes={scale=0.75, transform shape}},
		grid,
		y tick label style={
			/pgf/number format/.cd,
			fixed,
			fixed zerofill,
			precision=2,
			/tikz/.cd
		},
		xlabel=$N_c$,
		ylabel=MAE,
		ylabel shift = -4 pt,
		ymax=0.2,
		ymin=0.06,
		ytick={0.08,0.10,...,0.2},
		xtick={0,40,...,200},
		legend entries={$N_o=200$, $N_o=300$, $N_o=400$},
		]
		\addplot[color=cyan,mark=*,mark size=1.1,thick] table{results/mae_ray_200.txt};
		\addplot[color=orange,mark=triangle*,mark size=1.5,thick] table{results/mae_ray_300.txt};
		\addplot[color=purple,mark=square*,mark size=1.1,thick] table{results/mae_ray_400.txt};
		\end{axis}
		\end{tikzpicture}
	}\hfil    
	\subfloat[Gompertz distribution]{
		\begin{tikzpicture}[trim axis left, trim axis right]
		\begin{axis}
		[
		tiny,
		width=0.4\columnwidth,
		height=4.5cm,
		legend pos=north east,
		legend style={font=\scriptsize,nodes={scale=0.75, transform shape}},
		grid,
		y tick label style={
			/pgf/number format/.cd,
			fixed,
			fixed zerofill,
			precision=2,
			/tikz/.cd
		},
		xlabel=$N_c$,
		ylabel=MAE,
		ylabel shift = -4 pt,
		ymax=0.24,
		ymin=0.03,
		ytick={0.06,0.09,...,0.21},
		xtick={0,40,...,200},
		legend entries={$N_o=200$, $N_o=300$, $N_o=400$},
		]
		\addplot[color=cyan,mark=*,mark size=1.1,thick] table{results/mae_gom_200.txt};
		\addplot[color=orange,mark=triangle*,mark size=1.5,thick] table{results/mae_gom_300.txt};
		\addplot[color=purple,mark=square*,mark size=1.1,thick] table{results/mae_gom_400.txt};
		\end{axis}
		\end{tikzpicture}
	}
	\caption{\npglm's mean absolute error (MAE) vs the number of censored samples ($N_c$) for different number of observed samples ($N_o$).}
	\label{fig:syn-mae-c}
\end{figure}
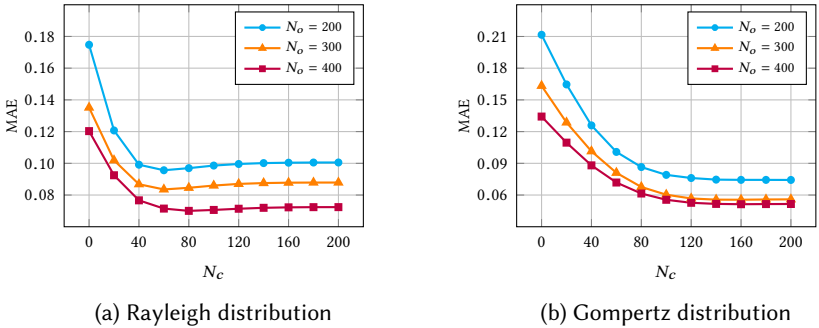

In another experiment, we investigated whether censored samples are informative or not. For this purpose, we fixed the number of observed samples $N_o$ and changed the number of censored samples from 0 to 200. We measured the MAE between the learned $\mb{w}$ and the ground truth for $N_o\in\{200,300,400\}$. The result is shown in Fig.~\ref{fig:syn-mae-c}. It clearly demonstrates that adding more censored samples causes the MAE to dwindle up to an extent, after which we get no substantial improvement. This threshold is dependent on the underlying distribution. In this case, for Rayleigh and Gompertz it is about 80 and 120, respectively.

\subsubsection{Running Time Analysis}    
\begin{figure}[t]
    \subfloat[Rayleigh distribution]{
        \begin{tikzpicture}[trim axis left, trim axis right]
        \begin{axis}
        [
        tiny,
        width=0.4\columnwidth,
        height=4.5cm,
        legend pos=south east,
        legend style={font=\scriptsize,nodes={scale=0.75, transform shape}},
        grid,
        xlabel=$N$,
        ylabel=$T\ \ (seconds)$,
        ylabel shift = -4 pt,
        xmode=log,
        ymode=log,
        legend entries={0\% censoring, 25\% censoring, 50\% censoring},
        ]
        \addplot[color=purple,mark=square*,mark size=1.1,thick] table{results/time_ray_5.txt};
        \addplot[color=cyan,mark=*,mark size=1.1,thick] table{results/time_ray_25.txt};
        \addplot[color=orange,mark=triangle*,mark size=1.5,thick] table{results/time_ray_50.txt};
        \end{axis}
        \end{tikzpicture}
    }\hfil
    \subfloat[Gompertz distribution]{
        \begin{tikzpicture}[trim axis left, trim axis right]
        \begin{axis}
        [
        tiny,
        width=0.4\columnwidth,
        height=4.5cm,
        legend pos=south east,
        legend style={font=\scriptsize,nodes={scale=0.75, transform shape}},
        grid,
        xlabel=$N$,
        ylabel=$T\ \ (seconds)$,
        ylabel shift = -4 pt,
        xmode=log,
        ymode=log,
        legend entries={0\% censoring, 25\% censoring, 50\% censoring},
        ]
        \addplot[color=purple,mark=square*,mark size=1.1,thick] table{results/time_gom_5.txt};
        \addplot[color=cyan,mark=*,mark size=1.1,thick] table{results/time_gom_25.txt};
        \addplot[color=orange,mark=triangle*,mark size=1.5,thick] table{results/time_gom_50.txt};
        \end{axis}
        \end{tikzpicture}
    }
    \caption{\npglm's average running time ($T$) measured in seconds vs the number of training samples ($N$) in $\log-\log$ scale for different censoring ratios.}
    \label{fig:syn-time-n}
\end{figure}
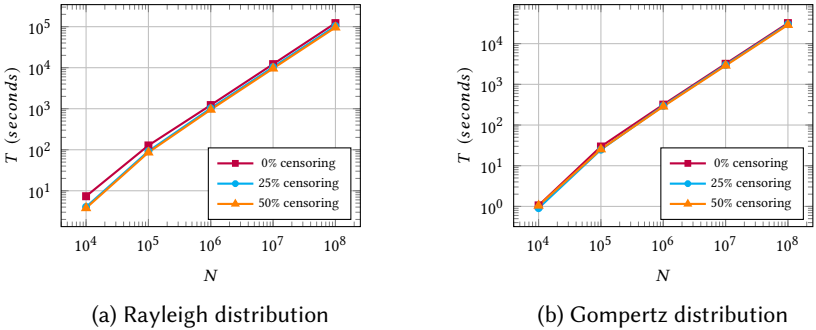

Finally, we assess the running time of \npglm's learning algorithm against the size of the training data when it becomes relatively large. To this end, we varied the number of samples from 10K to 100M and measured the average running time of the learning algorithm of \npglm on a single machine whose specification is reported in Table~\ref{table:pc}. Fig.~\ref*{fig:syn-time-n} depicts the result in log-log scale for Rayleigh and Gompertz distributions under different censoring ratios selected from the set $\{0.05, 0.25, 0.50\}$. It can be seen from the figure that the running time scales linearly with the number of training samples since the number of parameters to be inferred in \npglm as a non-parametric model depends on the size of the training data. The censoring ratio though negligible in scale can impact the running time of the algorithm, with more censoring ratio resulting in less running time. This is because higher censoring ratio reduces the observation window, which in turn reduces the number of parameters.

\begin{table}[t!]
	\centering
	\caption{PC Specification and Configuration}
	\label{table:pc}
	\begin{tabular} {c c}
		\toprule
		Operating System & Windows 10\\
		CPU & Intel Core i7 1.8 GHz\\
		RAM & 12 GB DDR III\\
		GPU & Nvidia GeForce GT 750\\
		Disk Type & SSD\\
		Programming Language & Python 3.6 \\
		\bottomrule 
	\end{tabular}
\end{table}

\section{Experiments on Real Data}\label{sec:results}

We apply \npglm with the proposed feature set on a number of real-world datasets to evaluate its effectiveness and compare its performance in predicting the relationship building time vis-\`a-vis state of the art models. 

\subsection{Datasets}
\subsubsection{DBLP}
We use the DBLP bibliographic citation network, provided by \cite{tang2008aminer}, which has both attributes of dynamicity and heterogeneity. The network contains four types of objects: authors, papers, venues, and terms. The network schema of this dataset is depicted in Fig.~\ref{fig:schema:dblp}. Each paper is associated with a publication date, with a granularity of one year. Based on the publication venue of the papers, we limited the original DBLP dataset to those papers that are published in venues relative to the theoretical computer science. This resulted in having about 16k authors and 37k papers published from 1969 to 2016 in 38 venues. 

\subsubsection{Delicious}
Another dynamic and heterogeneous dataset we use in our experiments is the Delicious bookmarking dataset from \cite{Cantador:RecSys2011}, with a network schema presented in Fig.~\ref{fig:schema:delicious}. It contains three types of objects, namely users, bookmarks, and tags, whose numbers are about 1.7k, 31k, and 22k, respectively. The dataset includes bookmarking timestamps from May 2006 to October 2010.

\subsubsection{MovieLens}
The third heterogeneous dataset with dynamic characteristics has been extracted from MovieLens personalized movie recommendation website, provided by \cite{harper2015}. The dataset comprises seven types of objects, that are users, movies, tags, genres, actors, directors, and countries, as illustrated by the network schema in the Fig.~\ref{fig:schema:movielens}. It contains about 1.4k users and 5.6k movies, with user-movie rating timestamps ranging from September 1997 to January 2009.

The demographic statistics of all datasets are presented in Table~\ref{table:dataset}.

\begin{table}[t]
    \centering
    \caption{Demographic Statistics of Real-World Datasets}
    \label{table:dataset}
    \footnotesize
    \begin{tabular} {l l l l r l r}
        \toprule
        Dataset & Time Span & Entity & \multicolumn{4}{c}{Count}\\
        \midrule 
        \multirow{4}{*}{DBLP} & \multirow{4}{*}{From 1969 to 2016}
        & \multirow{2}{*}{Nodes}
        & $Author$ & 15,929 & $Venue$ & 38 \\ 
        & & & $Paper$ & 37,077 & $Term$ & 12,028 \\ 
        \cmidrule{3-7}
        & & \multirow{2}{*}{Links}
        & write & 100,797 & publish & 42,872 \\ 
        & & & cite & 165,904 & mention & 284,156 \\ 
        
        \midrule 
        \multirow{4}{*}{Delicious} & \multirow{4}{*}{From May 2006 to Oct 2010}
        & \multirow{2}{*}{Nodes}
        & $User$ & 1,714 & $Tag$ & 21,956 \\ 
        & & & $Bookmark$ & 30,998 & & \\ 
        \cmidrule{3-7}
        & & \multirow{2}{*}{Links}
        & contact & 15,329 & has-tag & 437,594 \\ 
        & & & post & 437,594 & & \\ 
        
        \midrule 
        \multirow{7}{*}{MovieLens} & \multirow{7}{*}{From Sep 1997 to Jan 2009}
        & \multirow{4}{*}{Nodes}
        & $User$ & 1,421 & $Genre$ & 19 \\ 
        & & & $Movie$ & 5,660 & $Tag$ & 5,561 \\ 
        & & & $Actor$ & 6,176 & $Country$ & 63 \\ 
        & & & $Director$ & 2,401 & & \\ 
        \cmidrule{3-7}
        & & \multirow{3}{*}{Links}
        & rate & 855,599 & has-genre & 20,810 \\ 
        & & & play-in & 231,743 & has-tag & 47,958 \\ 
        & & & direct & 10,156 & produced-in & 10,198 \\ 
        \bottomrule 
    \end{tabular}
\end{table}

\subsection{Experiment Settings}
\subsubsection{Comparison Methods}
To challenge the performance of \npglm, we use a number of baselines introduced in the following:

\begin{itemize}
\item \emph{Generalized Linear Model (\textsc{Glm})}: This is the state-of-the-art method proposed in \cite{sun2012will}. We use the GLM-based framework with Exponential and Weibull distributions, denoted as \textsc{Exp-Glm} and \textsc{Wbl-Glm} used in \cite{sun2012will}.

\item \emph{Censored Regression Model (\textsc{Crm})}: This model, also called type II Tobit model, is designed to estimate linear relationships between variables when there is censoring in the dependent variable. In other words, it is an extension to the ordinary least squares linear regression for cencored data \cite{tobit}. The structural equation in this model is:
\[ t^*=\mb{w}^T\mb{x}+\epsilon \]
where $\epsilon$ is a normally distributed error term and $t^*$ is a latent variable which is observed within the observation window and censored otherwise. Accordingly, the observed $t$ is defined as:
\[ t=\begin{cases}
t^* & \text{if}\quad y=1 \\
\Omega & \text{if}\quad y=0\\
\end{cases} \]
The coefficient vector $\mb{w}$ is learned using maximum likelihood estimation (more details in \cite{amemiya1984tobit}).
\item \emph{Additive Regression Model (\textsc{Arm})}: This model is another regression method suggested by Aalen for censored data \cite{aalen1989linear}. Like \npglm, it specifies the intensity function, but instead of a multiplicative linear model, the Aalen's model is additive:
\[\lambda(t\mid \mb{x})=\sum_{i=0}^dw_i(t)x_i \]
The learning algorithm infers $\int_{0}^{t}w_i(t)dt$ instead of estimating individual $w_i$s. 
For more details about the learning algorithm, the reader can refer to \cite{hosmer2011applied}.

\end{itemize}

For all models, we consider the median of the distribution $f_T(t\mid\mb{x}_{test})$ as the predicted time for any test sample and then compare it to the ground truth time $t_{test}$.

To examine the effect of considering different feature extractors on the performance of the models, we use another dynamic feature extractor and a static one against the proposed LSTM Autoencoder:
\begin{itemize}
\item \emph{Exponential Smoothing}: This dynamic feature extractor previously used in \cite{hajibagheri2016leveraging} is an exponentially weighted moving average over the features extracted in all the snapshots of the network, which is calculated as:
\[\mb{f}^i=\begin{cases} 
\mb{x}^1, & \text{if}\quad i=1 \\
\alpha\mb{x}^i+(1-\alpha)\mb{f}^{i-1}, & \text{otherwise}
\end{cases}\]
where $\mb{f}^i$ is the smoothed feature after $i$th snapshot, $\mb{x}^i$ is the $i$th step of the dynamic meta-path-based time series, and $\alpha\in(0,1)$ is the smoothing factor. We then set $\mb{x}=\mb{f}^k$ as the final feature vector if we have $k$ snapshots in total.
\item \emph{Single Snapshot}: This static feature extractor considers the whole network as a single snapshot, neglecting its temporal dynamics. This feature extractor is equivalent to the one proposed in \cite{sun2012will}.
\end{itemize}

\subsubsection{Performance Measures}
We assess different methods using a number of evaluation metrics which are described in the following:
\begin{itemize}
\item Mean Absolute Error (MAE): This metric measures the expected absolute error between the predicted time values and the ground truth:
\[MAE(\mb{t},\hat{\mb{t}}) = \frac{1}{N}\sum_{i=1}^{N}\left|t_i-\hat{t}_i\right|\]
\item Mean Relative Error (MRE): This metric calculates the expected relative absolute error between the predicted time values and the ground truth:
\[MRE(\mb{t},\hat{\mb{t}}) = \frac{1}{N}\sum_{i=1}^{N}\left|\frac{t_i-\hat{t}_i}{t_i}\right|\]
\item Root Mean Squared Error (RMSE): This metric computes the root of the expected squared error between the predicted time values and the ground truth:
\[RMSE(\mb{t},\hat{\mb{t}}) = \sqrt{\frac{1}{N}\sum_{i=1}^{N}\left(t_i-\hat{t}_i\right)^2}\]
\item Mean Squared Logarithmic Error (MSLE): This measures the expected value of the squared logarithmic error between the predicted time values and the ground truth:
\[MSLE(\mb{t},\hat{\mb{t}}) = \frac{1}{N}\sum_{i=1}^{N}\left(\log{(1+t_i)}-log{(1+\hat{t}_i)}\right)^2\]
\item Median Absolute Error (MDAE): It is the median of the absolute errors between the predicted time values and the ground truth:
\[MDAE(\mb{t},\hat{\mb{t}}) = median(\left|t_1-\hat{t}_1\right|\dots\left|t_N-\hat{t}_N\right|)\]
\item Maximum Threshold Prediction Accuracy (ACC): This measures for what fraction of samples, a model have a lower absolute error than a given threshold:
\[ACC(\mb{t},\hat{\mb{t}})=\frac{1}{N}\sum_{i=1}^{N}\mb{1}\left(\left|t_i-\hat{t}_i\right| < threshold\right)\]
\item Concordance Index (CI): This metric is one of the most widely used performance measures for survival models that estimates how good a model performs at ranking predicted times \cite{harrell1982evaluating}. It can be seen as the fraction of all the sample pairs whose predicted timestamps are correctly ordered among all samples that can be ordered, and is considered as the generalization of the Area Under Receiver Operating Characteristic Curve (AUC) when we are dealing with censored data \cite{steck2008ranking}.
\end{itemize}

\subsubsection{Experiment Setup}

For DBLP dataset, we confine the data samples to those authors who have published more than 5 papers in the feature extraction window of each experiment. Following the triple building blocks described for feature extraction in Section~\ref{sec:features}, and using the similarity meta-paths in Table~\ref{table:meta}, we start the feature extraction process with 19 feature meta-paths. In all experiments, the author citation relation ($A\rightarrow P\rightarrow P\leftarrow A$) is chosen as the target relation. 
For the case of the Delicious dataset, we select user-user relation ($U\leftrightarrow U$) as the target relation, and design 6 feature meta-paths via the similarity meta-paths in Table~\ref{table:meta}.
Regarding the MovieLens dataset, we limit the actor list to the top three for each movie. To imply a notion of ``like'' relation between user and movie, we only consider ratings above 4 in the scale of 5. For this dataset, the target relation is set to user rate movie ($U\rightarrow M$), based on which, we design 11 final meta-paths. For the sake of convenience, we convert the scale of time differences from timestamp to month in Delicious and MovieLens datasets.

Except for parameter settings analysis (subsection~\ref{sec:param-analysis}) where we will analyze the effect of different parameters on the performance of different models, in the rest of the experiments in this section we set the length of the observation window $\Omega$ to 6 for all three datasets. For DBLP dataset, the number of snapshots $k$ is set to 6, while for the other two datasets we set $k=12$. We also fix the time difference between network snapshots $\Delta$ to 1 in all cases. These settings lead to having a feature extraction window of size $\Phi=6$ years for DBLP, and $\Phi=12$ months for Delicious and MovieLens. Accordingly, the number of labeled instances for DBLP, Delicious, and MovieLens are about 3.4K, 3.9K, and 7.8K, respectively. About half of the labeled samples are censored ones, which are picked uniformly at random among all the possible candidates.

We implemented the LSTM autoencoder using Keras deep learning library \cite{chollet2015keras}. We used mean square error loss function, linear activation function, and Adadelta optimizer \cite{zeiler2012adadelta} with default parameters. For all datasets, we set the dimension of the encoded feature as twice as the input dimension and trained the autoencoder in 50 epochs. For exponential smoothing feature extractor, the smoothing factor $\alpha$ were tuned to maximize the performance on the training dataset. For Np-Glm, the data samples were ordered according to their corresponding time variables, as the model needs the samples sorted by their recorded time. We use 5-fold cross-validation and report the average results for all the experiments in this section.

\subsection{Experiment Results}
In the rest of this section, we first assess how well different methods perform over various datasets and compare their performance based on different measures. Next, we discuss the efficiency of our proposed method by measuring and comparing its running time against the other baselines. Finally, we analyze the effect of different parameters and problem configurations on the performance of competitive methods.

\begin{table}[t]
    \centering
    \caption{Comprehensive Performance Comparison of Different Methods}
    \label{table:results}
    \scriptsize
    \begin{tabu} to \columnwidth {X[c] c X[l] X[r] X[r] X[r] X[r] X[r] X[r]}
        \toprule
        Dataset & Feature &
        Model &  MAE &   MRE &   RMSE &   MSLE &   MDAE &  CI \\
        \midrule
        \multirow{15}{*}[-3em]{\rotatebox{90}{DBLP}}
        & \multirow{6}{*}{\shortstack{LSTM Autoencoder\\(Dynamic)}}
        & \npglm  &  $\bm{1.99}$ &  $\bm{0.95}$ &   $\bm{2.43}$ &   $\bm{0.30}$ &  $\bm{1.73}$ & $\bm{0.62}$ \\
        & & \textsc{Wbl-Glm} &  2.33 &  1.10 &   2.85 &   0.36 &   2.08 & 0.58 \\
        & & \textsc{Exp-Glm} &  3.11 &  1.39 &   3.88 &   0.52 &   2.58 & 0.50 \\
        & & \textsc{Crm} & 3.08 & 1.06 & 3.32 & 2.04 & 2.98 & 0.37 \\
        & & \textsc{Arm} & 2.95 & 1.33 & 4.48 & 0.48 & 1.48 & 0.56 \\
        
        \cmidrule{2-9}                                                                            
        & \multirow{6}{*}{\shortstack{Exp. Smoothing\\(Dynamic)}}
        & \npglm               &  2.15 &  1.07  &  2.54  &  0.32  &  1.98 & 0.53 \\
        & & \textsc{Wbl-Glm}     &  2.50 &  1.22 &   2.89  &  0.38  &  2.46 & 0.58 \\
        & & \textsc{Exp-Glm}     &  3.20 &  1.49  &  3.73  &  0.51  &  3.06&  0.45 \\
        & & \textsc{Crm} & 2.55 & 0.97 & 3.05 & 1.58 & 2.11 & 0.55 \\
        & & \textsc{Arm} & 6.75 & 2.83 & 7.86 & 1.17 & 6.39 & 0.60 \\

        \cmidrule{2-9}                                                                            
        & \multirow{6}{*}{\shortstack{Single Snapshot\\(Static)}}                                                  
        & \npglm               &  2.76 &  1.35 &   3.07 &   0.44 &   2.88 & 0.50 \\
        & & \textsc{Wbl-Glm}     &  2.81 &  1.38 &   3.16 &   0.45 &   2.88 & 0.48 \\
        & & \textsc{Exp-Glm}     &  3.28 &  1.57 &   3.70 &   0.53 &   3.30 & 0.14 \\
        & & \textsc{Crm} & 2.96 & 1.03 & 3.21 & 1.73 & 2.97 & 0.38 \\
        & & \textsc{Arm} & 3.89 & 1.76 & 5.45 & 0.66 & 2.11 & 0.46 \\
        
        
        \midrule
        \multirow{15}{*}[-3em]{\rotatebox{90}{Delicious}}
        & \multirow{6}{*}{\shortstack{LSTM Autoencoder\\(Dynamic)}}
        & \npglm  &  $\bm{2.10}$ &  $\bm{1.20}$ &   $\bm{2.55}$ &   $\bm{0.35}$ &   $\bm{2.05}$ & $\bm{0.70}$ \\
        & & \textsc{Wbl-Glm} &  2.37 &  1.31 &   2.89 &   0.40 &   2.16 & 0.57 \\
        & & \textsc{Exp-Glm} &  3.21 &  1.58 &   3.84 &   0.54 &   2.89 & 0.55 \\
        & & \textsc{Crm} & 6.38 & 3.10 & 6.55 & 1.33 & 6.87 & 0.43 \\
        & & \textsc{Arm} & 5.20 & 2.56 & 6.23 & 0.86 & 4.99 & 0.52 \\
        
        \cmidrule{2-9}                 
        & \multirow{6}{*}{\shortstack{Exp. Smoothing\\(Dynamic)}}                                                  
        & \npglm               &  2.25  & 1.36  &  2.74  &  0.40  &  2.11 & 0.66 \\
        & & \textsc{Wbl-Glm}     &  2.61  & 1.64  &  3.20 &   0.47   & 2.17 & 0.56 \\
        & & \textsc{Exp-Glm}     &  3.52  & 1.99  &  4.54  &  0.62  &  3.20  &0.39 \\
        & & \textsc{Crm} & 3.28 & 3.69 & 3.84 & 2.07 & 2.88 & 0.43 \\
        & & \textsc{Arm} & 6.36 & 3.24 & 7.80 & 1.09 & 6.72 & 0.56 \\
        
        \cmidrule{2-9}                 
        & \multirow{6}{*}{\shortstack{Single Snapshot\\(Static)}}                                                  
        & \npglm               &  2.33 &  1.46 &   2.80 &   0.41 &   2.17 & 0.61 \\
        & & \textsc{Wbl-Glm}     &  2.65 &  1.62 &   3.23 &   0.47 &   2.26 & 0.43 \\
        & & \textsc{Exp-Glm}     &  3.35 &  1.91 &   4.17 &   0.59 &   2.75 & 0.35 \\
        & & \textsc{Crm} & 3.06 & 2.05 & 3.47 & 1.53 & 2.84 & 0.38 \\
        & & \textsc{Arm} & 5.79 & 2.76 & 6.69 & 1.16 & 5.89 & 0.37 \\
        
        \midrule
        \multirow{15}{*}[-3em]{\rotatebox{90}{MovieLens}}
        & \multirow{6}{*}{\shortstack{LSTM Autoencoder\\(Dynamic)}}
        & \npglm  &  $\bm{2.48}$ &  $\bm{3.08}$ &   $\bm{3.04}$ &   $\bm{0.55}$ &  $\bm{2.14}$ & $\bm{0.70}$ \\
        & & \textsc{Wbl-Glm} &  3.06 &  3.61 &   3.79 &   0.65 &   2.60 & 0.56 \\
        & & \textsc{Exp-Glm} &  3.79 &  2.70 &   4.60 &   0.78 &   3.48 & 0.45 \\
        & & \textsc{Crm} & 3.07 & 3.47 & 3.74 & 2.02 & 2.51 & 0.40 \\
        & & \textsc{Arm} & 5.53 & 5.63 & 7.41 & 1.12 & 3.80 & 0.53 \\
        
        \cmidrule{2-9}                                                                            
        & \multirow{6}{*}{\shortstack{Exp. Smoothing\\(Dynamic)}}                                                  
        & \npglm               &  2.69 &  3.35  &  3.18  &  0.59  &  2.61  &0.66 \\
        & & \textsc{Wbl-Glm}     &  3.09&   3.62 &   3.59 &   0.66 &   2.95 & 0.52 \\
        & & \textsc{Exp-Glm}     &  3.52 &  2.86  &  4.05  &  0.74  &  3.26 & 0.43 \\
        & & \textsc{Crm} & 3.18 & 3.37 & 3.68 & 1.90 & 2.56 & 0.48 \\
        & & \textsc{Arm} & 9.39 & 8.60 & 10.06 & 1.83 & 9.26 & 0.52 \\
        
        \cmidrule{2-9}                                                                            
        & \multirow{6}{*}{\shortstack{Single Snapshot\\(Static)}}                                                  
        & \npglm               &  2.92 &  3.44 &   3.45 &   0.67 &   3.36 & 0.50 \\
        & & \textsc{Wbl-Glm}     &  2.99 &  3.52 &   3.51 &   0.69 &   3.37 & 0.49 \\
        & & \textsc{Exp-Glm}     &  3.42 &  2.89 &   3.86 &   0.78 &   3.82 & 0.49 \\
        & & \textsc{Crm} & 3.14 & 3.48 & 3.63 & 2.20 & 3.55 & 0.35 \\
        & & \textsc{Arm} & 5.71 & 5.50 & 7.30 & 1.23 & 5.18 & 0.47 \\
        
        \bottomrule
    \end{tabu}
\end{table}

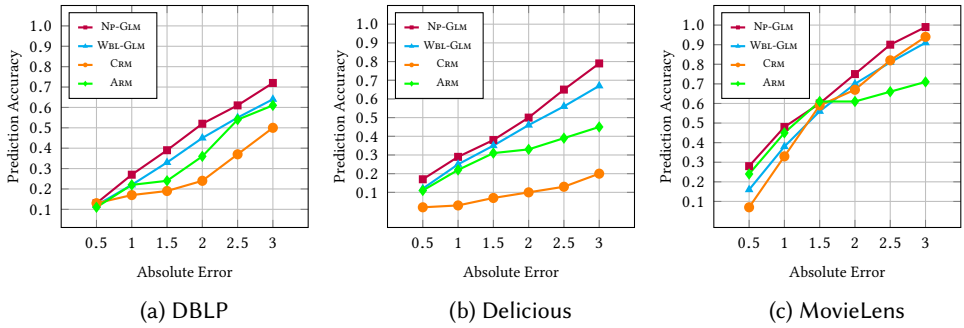
\begin{figure*}[t]
    \centering
    \subfloat[DBLP]{
        \begin{tikzpicture}[trim axis left, trim axis right]
        \begin{axis}
        [
        tiny,
        width=0.35\columnwidth,
        height=4.5cm,
        legend pos=north west,
        legend image post style={scale=0.4},
        legend style={font=\tiny,nodes={scale=0.75, transform shape}},
        grid,
        y tick label style={
            /pgf/number format/.cd,
            fixed,
            fixed zerofill,
            precision=1,
            /tikz/.cd
        },
        xlabel=Absolute Error,
        ylabel=Prediction Accuracy,
        ylabel shift = -4 pt,
        ymax=1.1,
        xmin=0,
        xmax=3.5,
        ytick={0.1,0.2,...,0.9,1.0},
        xtick={0.5,1.0,...,3},
        legend entries={\npglm, \textsc{Wbl-Glm}, \textsc{Crm}, \textsc{Arm}},
        ]
        \addplot[color=purple,mark=square*,mark size=1.1,thick] table{results/db_np.txt};
        \addplot[color=cyan,mark=triangle*,mark size=1.1,thick] table{results/db_wbl.txt};
        \addplot[color=orange,mark=*,mark size=1.5,thick] table{results/db_tob.txt};
        \addplot[color=green,mark=diamond*,mark size=1.5,thick] table{results/db_aam.txt};
        \end{axis}
        \end{tikzpicture}
    }
    \hfil
    \subfloat[Delicious]{
        \begin{tikzpicture}[trim axis left, trim axis right]
        \begin{axis}
        [
        tiny,
        width=0.35\columnwidth,
        height=4.5cm,
        legend pos=north west,
        legend image post style={scale=0.4},
        legend style={font=\tiny,nodes={scale=0.75, transform shape}},
        grid,
        y tick label style={
            /pgf/number format/.cd,
            fixed,
            fixed zerofill,
            precision=1,
            /tikz/.cd
        },
        xlabel=Absolute Error,
        ylabel=Prediction Accuracy,
        ylabel shift = -4 pt,
        ymax=1.1,
        xmin=0,
        xmax=3.5,
        ytick={0.1,0.2,...,0.9,1.0},
        xtick={0.5,1.0,...,3},
        legend entries={\npglm, \textsc{Wbl-Glm}, \textsc{Crm}, \textsc{Arm}},
        ]
        \addplot[color=purple,mark=square*,mark size=1.1,thick] table{results/dl_np.txt};
        \addplot[color=cyan,mark=triangle*,mark size=1.1,thick] table{results/dl_wbl.txt};
        \addplot[color=orange,mark=*,mark size=1.5,thick] table{results/dl_tob.txt};
        \addplot[color=green,mark=diamond*,mark size=1.5,thick] table{results/dl_aam.txt};
        \end{axis}
        \end{tikzpicture}
    }
    \hfil
    \subfloat[MovieLens\label{fig:real:movielens}]{
        \begin{tikzpicture}[trim axis left, trim axis right]
        \begin{axis}
        [
        tiny,
        width=0.35\columnwidth,
        height=4.5cm,
        legend pos=north west,
        legend image post style={scale=0.4},
        legend style={font=\tiny,nodes={scale=0.75, transform shape}},
        grid,
        y tick label style={
            /pgf/number format/.cd,
            fixed,
            fixed zerofill,
            precision=1,
            /tikz/.cd
        },
        xlabel=Absolute Error,
        ylabel=Prediction Accuracy,
        ylabel shift = -4 pt,
        ymax=1.1,
        xmin=0,
        xmax=3.5,
        ytick={0.1,0.2,...,0.9,1.0},
        xtick={0.5,1.0,...,3},
        legend entries={\npglm, \textsc{Wbl-Glm}, \textsc{Crm}, \textsc{Arm}},
        ]
        \addplot[color=purple,mark=square*,mark size=1.1,thick] table{results/mv_np.txt};
        \addplot[color=cyan,mark=triangle*,mark size=1.1,thick] table{results/mv_wbl.txt};
        \addplot[color=orange,mark=*,mark size=1.5,thick] table{results/mv_tob.txt};
        \addplot[color=green,mark=diamond*,mark size=1.5,thick] table{results/mv_aam.txt};
        \end{axis}
        \end{tikzpicture}
    }
    \caption{Prediction accuracy of different methods vs the maximum tolerated absolute error on different datasets.}
    \label{fig:real}
\end{figure*}

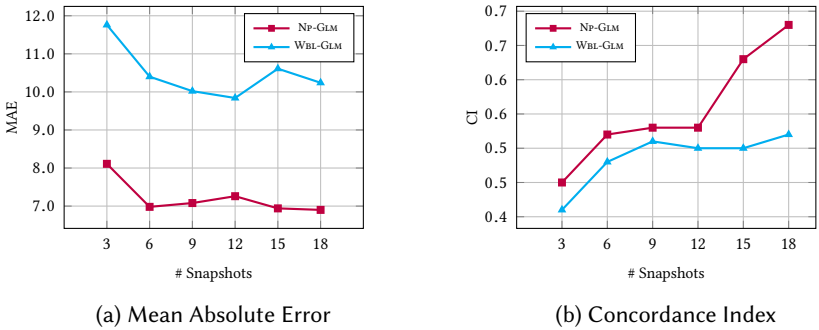
\begin{figure}[t]
    \centering
    \subfloat[Mean Absolute Error]{
        \begin{tikzpicture}[trim axis left, trim axis right]
        \begin{axis}
        [
        tiny,
        width=0.4\columnwidth,
        height=4.5cm,
        legend pos=north east,
        legend style={font=\tiny,nodes={scale=0.75, transform shape}},
        grid,
        y tick label style={
            /pgf/number format/.cd,
            fixed,
            fixed zerofill,
            precision=1,
            /tikz/.cd
        },
        xlabel=\# Snapshots,
        ylabel=MAE,
        ylabel shift = -4 pt,
        xmin=0,
        xmax=21,
        xtick={3,6,...,18},
        legend entries={\npglm, \textsc{Wbl-Glm}},
        ]
        \addplot[color=purple,mark=square*,mark size=1.1,thick] table{results/dl_snap_mae_np.txt};
        \addplot[color=cyan,mark=triangle*,mark size=1.1,thick] table{results/dl_snap_mae_wbl.txt};
        \end{axis}
        \end{tikzpicture}
    }
    \hfil
    \subfloat[Concordance Index]{
        \begin{tikzpicture}[trim axis left, trim axis right]
        \begin{axis}
        [
        tiny,
        width=0.4\columnwidth,
        height=4.5cm,
        legend pos=north west,
        legend style={font=\tiny,nodes={scale=0.75, transform shape}},
        grid,
        y tick label style={
            /pgf/number format/.cd,
            fixed,
            fixed zerofill,
            precision=1,
            /tikz/.cd
        },
        xlabel=\# Snapshots,
        ylabel=CI,
        ylabel shift = -4 pt,
        xmin=0,
        xtick={3,6,...,18},
        legend entries={\npglm, \textsc{Wbl-Glm}},
        ]
        \addplot[color=purple,mark=square*,mark size=1.1,thick] table{results/dl_snap_ci_np.txt};
        \addplot[color=cyan,mark=triangle*,mark size=1.1,thick] table{results/dl_snap_ci_wbl.txt};
        \end{axis}
        \end{tikzpicture}
    }
    \caption{Effect of choosing different number of snapshots on performance of different methods using Delicious dataset.}
    \label{fig:snaps:delicious}
\end{figure}

\begin{figure}[t]
    \centering
    \subfloat[Mean Absolute Error]{
        \begin{tikzpicture}[trim axis left, trim axis right]
        \begin{axis}
        [
        tiny,
        width=0.4\columnwidth,
        height=4.5cm,
        legend pos=north east,
        legend style={font=\tiny,nodes={scale=0.75, transform shape}},
        grid,
        y tick label style={
            /pgf/number format/.cd,
            fixed,
            fixed zerofill,
            precision=1,
            /tikz/.cd
        },
        xlabel=\# Snapshots,
        ylabel=MAE,
        ylabel shift = -4 pt,
        xmin=0,
        xtick={3,6,...,18},
        legend entries={\npglm, \textsc{Wbl-Glm}},
        ]
        \addplot[color=purple,mark=square*,mark size=1.1,thick] table{results/mv_snap_mae_np.txt};
        \addplot[color=cyan,mark=triangle*,mark size=1.1,thick] table{results/mv_snap_mae_wbl.txt};
        \end{axis}
        \end{tikzpicture}
    }
    \hfil
    \subfloat[Concordance Index]{
        \begin{tikzpicture}[trim axis left, trim axis right]
        \begin{axis}
        [
        tiny,
        width=0.4\columnwidth,
        height=4.5cm,
        legend pos=north west,
        legend style={font=\tiny,nodes={scale=0.75, transform shape}},
        grid,
        y tick label style={
            /pgf/number format/.cd,
            fixed,
            fixed zerofill,
            precision=1,
            /tikz/.cd
        },
        xlabel=\# Snapshots,
        ylabel=CI,
        ylabel shift = -4 pt,
        ymax=0.9,ymin=0.2,
        xmin=0,
        xtick={3,6,...,18},
        legend entries={\npglm, \textsc{Wbl-Glm}},
        ]
        \addplot[color=purple,mark=square*,mark size=1.1,thick] table{results/mv_snap_ci_np.txt};
        \addplot[color=cyan,mark=triangle*,mark size=1.1,thick] table{results/mv_snap_ci_wbl.txt};
        \end{axis}
        \end{tikzpicture}
    }
    \caption{Effect of choosing different number of snapshots on performance of different methods using MovieLens dataset.}
    \label{fig:snaps:MovieLens}
\end{figure}
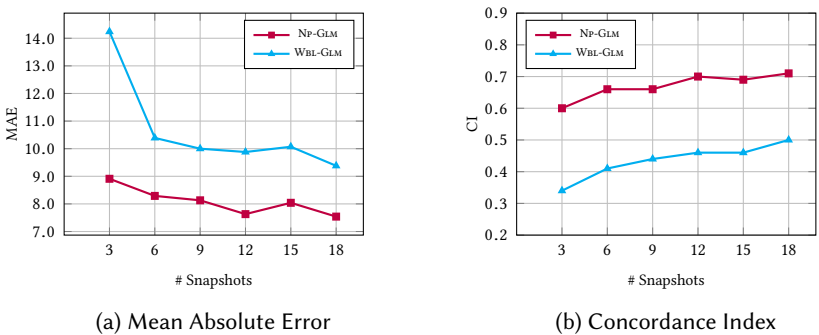

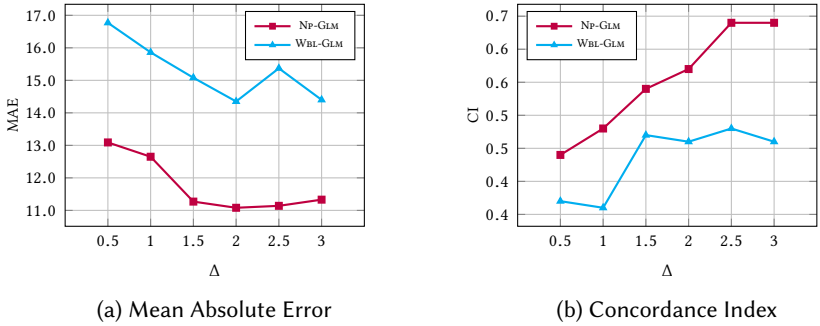
\begin{figure}[t]
    \centering
    \subfloat[Mean Absolute Error]{
        \begin{tikzpicture}[trim axis left, trim axis right]
        \begin{axis}
        [
        tiny,
        width=0.4\columnwidth,
        height=4.5cm,
        legend pos=north east,
        legend style={font=\tiny,nodes={scale=0.75, transform shape}},
        grid,
        y tick label style={
            /pgf/number format/.cd,
            fixed,
            fixed zerofill,
            precision=1,
            /tikz/.cd
        },
        xlabel=$\Delta$,
        ylabel=MAE,
        ylabel shift = -4 pt,
        xmin=0,
        xmax=3.5,
        xtick={0.5,1.0,...,3},
        legend entries={\npglm, \textsc{Wbl-Glm}},
        ]
        \addplot[color=purple,mark=square*,mark size=1.1,thick] table{results/delta_mae_np.txt};
        \addplot[color=cyan,mark=triangle*,mark size=1.1,thick] table{results/delta_mae_wbl.txt};
        \end{axis}
        \end{tikzpicture}
    }
    \hfil
    \subfloat[Concordance Index]{
        \begin{tikzpicture}[trim axis left, trim axis right]
        \begin{axis}
        [
        tiny,
        width=0.4\columnwidth,
        height=4.5cm,
        legend pos=north west,
        legend style={font=\tiny,nodes={scale=0.75, transform shape}},
        grid,
        y tick label style={
            /pgf/number format/.cd,
            fixed,
            fixed zerofill,
            precision=1,
            /tikz/.cd
        },
        xlabel=$\Delta$,
        ylabel=CI,
        ylabel shift = -4 pt,
        xmin=0,
        xmax=3.5,
        xtick={0.5,1.0,...,3},
        legend entries={\npglm, \textsc{Wbl-Glm}},
        ]
        \addplot[color=purple,mark=square*,mark size=1.1,thick] table{results/delta_ci_np.txt};
        \addplot[color=cyan,mark=triangle*,mark size=1.1,thick] table{results/delta_ci_wbl.txt};
        \end{axis}
        \end{tikzpicture}
    }
    \caption{Effect of choosing different values for $\Delta$ on performance of different methods using Delicious dataset.}
    \label{fig:delta:delicious}
\end{figure}

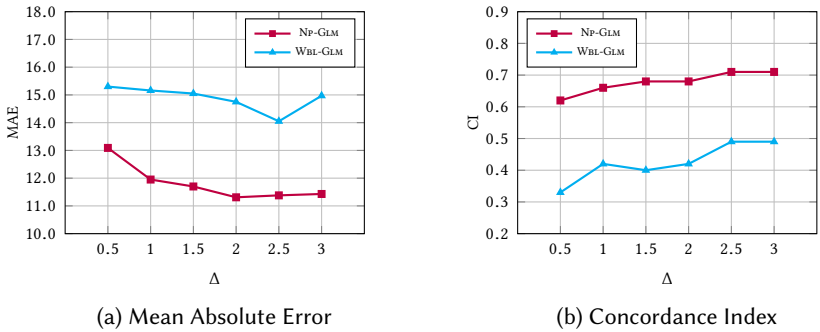
\begin{figure}[t]
    \centering
    \subfloat[Mean Absolute Error]{
        \begin{tikzpicture}[trim axis left, trim axis right]
        \begin{axis}
        [
        tiny,
        width=0.4\columnwidth,
        height=4.5cm,
        legend pos=north east,
        legend style={font=\tiny,nodes={scale=0.75, transform shape}},
        grid,
        y tick label style={
            /pgf/number format/.cd,
            fixed,
            fixed zerofill,
            precision=1,
            /tikz/.cd
        },
        xlabel=$\Delta$,
        ylabel=MAE,
        ylabel shift = -4 pt,
        ymax=18,ymin=10,
        xmin=0,
        xmax=3.5,
        xtick={0.5,1.0,...,3},
        legend entries={\npglm, \textsc{Wbl-Glm}},
        ]
        \addplot[color=purple,mark=square*,mark size=1.1,thick] table{results/mv_delta_mae_np.txt};
        \addplot[color=cyan,mark=triangle*,mark size=1.1,thick] table{results/mv_delta_mae_wbl.txt};
        \end{axis}
        \end{tikzpicture}
    }
    \hfil
    \subfloat[Concordance Index]{
        \begin{tikzpicture}[trim axis left, trim axis right]
        \begin{axis}
        [
        tiny,
        width=0.4\columnwidth,
        height=4.5cm,
        legend pos=north west,
        legend style={font=\tiny,nodes={scale=0.75, transform shape}},
        grid,
        y tick label style={
            /pgf/number format/.cd,
            fixed,
            fixed zerofill,
            precision=1,
            /tikz/.cd
        },
        xlabel=$\Delta$,
        ylabel=CI,
        ylabel shift = -4 pt,
        ymax=0.9,ymin=0.2,
        xmin=0,
        xmax=3.5,
        xtick={0.5,1.0,...,3},
        legend entries={\npglm, \textsc{Wbl-Glm}},
        ]
        \addplot[color=purple,mark=square*,mark size=1.1,thick] table{results/mv_delta_ci_np.txt};
        \addplot[color=cyan,mark=triangle*,mark size=1.1,thick] table{results/mv_delta_ci_wbl.txt};
        \end{axis}
        \end{tikzpicture}
    }
    \caption{Effect of choosing different values for $\Delta$ on performance of different methods using MovieLens dataset.}
    \label{fig:delta:MovieLens}
\end{figure}

\subsubsection{Comparative Performance Analysis}
In the first set of experiments, we evaluate the prediction power of different models combined with different feature extractors on DBLP, Delicious and MovieLens datasets. MAE, MRE, RMSE, MSLE, MDAE, and CI of all models using both dynamic and static feature sets has been shown in Table~\ref{table:results}. We see that in all three networks, \npglm with the LSTM Autoencoder feature set is superior to the other methods under all performance measures. For instance, our model \npglm can obtain an MAE of 1.99 for DBLP dataset, which is 15\% lower than the MAE obtained by its closest competitor, \textsc{Wbl-Glm}. As of CI, \npglm achieves 0.62 on DBLP, which is 7\% better than \textsc{Wbl-Glm}. On Delicious dataset, \npglm improves MAE and CI by 11\% and 23\%, respectively, relative to \textsc{Wbl-Glm}. Similarly, \npglm reduces MAE by 19\% and increases CI by 25\%. Comparable results hold for other performance measures as well. Accordingly, \textsc{Wbl-Glm}, which has two degrees of freedom, has shown a better performance compared to other models. That is while \npglm, as a non-parametric model with highly tunable shape, outperforms all the other ``less-flexible'' models by learning the true distribution of the data.

Moreover, it is evident from Table~\ref{table:results} that using the dynamic features learned with the LSTM autoencoder has boosted the performance of all models over different datasets, and has outperformed the other feature extractors. Based on the results presented in Table~\ref{table:results}, the alternative dynamic feature extractor, exponential smoothing, has performed better than the static single snapshot feature extractor, yet not better than the proposed LSTM Autoencoder. Comparing the LSTM Autoencoder with exponential smoothing feature extractor, over the DBLP dataset, the proposed feature extractor has achieved 7\% less MAE and 17\% more CI with \npglm. Over Delicious, \npglm with LSTM-based features reduces MAE by about 7\% and improves CI by 7\%. Finally, on the MovieLens dataset, combining LSTM Autoencoder and \npglm leads to an improvement of 8\% and 7\% under MAE and CI, respectively. The other models behave more or less similarly when they are combined with different feature extractors. This result clearly demonstrates that our feature extraction framework is performing well on capturing the temporal dynamics of the networks.

In the next experiment, we investigated the performance of different models using the LSTM autoencoder feature extraction framework under maximum threshold prediction accuracy. To evaluate the prediction accuracy of a model, we record the fraction of test samples for which the difference between their true times and predicted ones are lower than a given threshold, called \emph{tolerated error}. The results are plotted in Fig.~\ref{fig:real} where we varied the tolerated error in the range $\{0.5, 1.0, \dots, 3.0\}$. We can see from the figure that \npglm and \textsc{Wbl-Glm} perform comparably, yet \npglm outperforms \textsc{Wbl-Glm} in all cases. For example on MovieLens dataset (Fig.~\ref{fig:real:movielens}), \npglm can predict the relationship building time of all the test samples with 100\% accuracy by an error of 3 months, whereas for \textsc{Wbl-Glm}, this is reduced to 90\%. Similarly, on the Delicious dataset, \npglm with 3 months of tolerated error achieves around 80\% accuracy, which is about 12\% more than \textsc{Wbl-Glm}.

\subsubsection{Efficiency Analysis}
In this part, we analyze and compare the running time of \npglm and \textsc{Wbl-Glm} models utilizing different feature extractors, namely LSTM autoencoder, exponential smoothing, and single snapshot. All the algorithms were implemented in Python and were run on a Windows 10 PC with Intel Core i7 1.8 GHz CPU and 12GB of RAM. The full specification of the host machine is reported in Table~\ref{table:pc}. We measured the running time of all the methods during a complete training and test procedure, including feature extraction, learning, and inference. For exponential smoothing feature extractor, we included the time required for tuning the smoothing factor $\alpha$ using a separate validation set, while for LSTM based framework the training time of the autoencoder is counted toward total running time. Table~\ref{table:efficiency} presents the results over each of the DBLP, Delicious, and MovieLens datasets. Since a considerable amount of running time is spent on feature extraction, dynamic feature extraction frameworks require more time to process the network data as opposed to static single snapshot feature extractors. However, the proposed LSTM autoencoder performs comparably to exponential smoothing in terms of running time. Even though LSTM autoencoder is a bit slower than the other dynamic feature extractors, it demonstrates higher prediction performance compared to models utilizing exponential smoothing. For example, on MovieLens network with more than 20K nodes and 1 million links, \npglm with LSTM autoencoder requires less than four minutes to process the whole network, extract features and learn from about 6K labeled samples, and perform prediction for about 2K instances on a typical PC.

\begin{table}[t]
    \centering
    \caption{Comparison of Computational Time Measured in Seconds}
    \label{table:efficiency}
    \footnotesize
    \begin{tabu} to \columnwidth {c c c c c}
        \toprule
        \multirow{2}{*}{Dataset} & \multirow{2}{*}{Model} &
        \multicolumn{3}{c}{Feature Extractor} \\
        \cmidrule{3-5}
        & & Single Snapshot & Exp. Smoothing & LSTM Autoencoder\\
        
        \midrule
        \multirow{2}{*}{DBLP} & \npglm  &  35.92 &  98.35 &   93.59\\
                              & \textsc{Wbl-Glm} &  36.01 &  74.34 &   79.83 \\
        
        \midrule
        \multirow{2}{*}{Delicious} & \npglm  &  2.01 & 110.67 &   128.43\\
        & \textsc{Wbl-Glm} &  1.89 &  97.44 &   123.53 \\
        
        \midrule
        \multirow{2}{*}{MovieLens} & \npglm  &  19.60 &  177.015 &   232.77\\
        & \textsc{Wbl-Glm} &  19.70 &  154.86 &   213.7\\
        \bottomrule
    \end{tabu}
\end{table}

\subsubsection{Parameter Setting Analysis}\label{sec:param-analysis}
The performance of different models is influenced by two parameters, the number of snapshots $k$, and the time difference between snapshots $\Delta$, as these parameters determine the length of the feature extraction window $\Phi$. In this set of experiments, we investigate how these parameters affect the performance of our model \npglm and its closest competitor \textsc{Wbl-Glm} over Delicious and MovieLens datasets using the proposed LSTM based feature extraction framework. 

Firstly, The effect of increasing the number of snapshots on achieved MAE and CI by \npglm and \textsc{Wbl-Glm} over Delicious and MovieLens datasets is illustrated in Fig.~\ref{fig:snaps:delicious} and Fig.~\ref{fig:snaps:MovieLens}, respectively. For both datasets, we set $\Delta=1.5$ and $\Omega=18$ and varied the number of snapshots in the range of 3 to 18. As we can see in both figures, increasing the number of snapshots results in lower prediction error and higher accuracy. This is due to the fact that as the number of snapshots grows, a longer history of the network is taken into account. Therefore, different models can benefit from more information about the temporal dynamics of the network given to them through the extracted feature vector.

Finally, the impact of choosing different values for $\Delta$ is analyzed on the performance of \npglm and \textsc{Wbl-Glm} in terms of MAE and CI. The results for Delicious and MovieLens datasets are depicted in Fig.~\ref{fig:delta:delicious} and Fig.~\ref{fig:delta:MovieLens}, respectively. In this experiment, the number of snapshots and observation window length are accordingly set to 6 and 24. Different values of $\Delta$ are selected from the set $\{0.5,1.0,\dots,3.0\}$. As illustrated in both figures, by increasing  $\Delta$ up to an extent, we witness that the performance of models improves gradually. That is because increasing the value of $\Delta$ leads to a wider feature extraction window. However, since the number of snapshots is constant, we see no performance improvement when the value of $\Delta$ becomes greater than a certain threshold. This is due to the fact that short-term temporal evolution of the network will be ignored when the value of $\Delta$ becomes too wide.

\section{Related Works}\label{sec:related}
\newcommand{\etal}{\textit{et~al}.}

The problem of link prediction has been studied extensively in recent years and many approaches have been proposed to solve this problem \cite{wang2015link,wang2014review}.
Previous work on time-aware link prediction has mostly considered temporality in analyzing the long-term network trend over time \cite{dhote2013survey}. Authors in \cite{potgieter2009temporality} have shown that temporal metrics are an extremely valuable new contribution to link prediction, and should be used in future applications. 
Dunlavy \etal{} focused on the problem of periodic temporal link prediction \cite{dunlavy2011temporal}. They concentrated on bipartite graphs that evolve over time and also considered a weighted matrix that contained multilayer data and tensor-based methods for predicting future links.
Oyama \etal{} solved the problem of cross-temporal link prediction, in which the links among nodes in different time frames are inferred \cite{oyama2011cross}. They mapped data objects in different time frames into a common low-dimensional latent feature space and identified the links on the basis of the distance between the data objects.
{\"O}zcan \etal{} proposed a novel link prediction method for evolving networks based on NARX neural network \cite{ozcan2016temporal}. They take the correlation between the quasi-local similarity measures and temporal evolutions of link occurrences information into account by using NARX for multivariate time series forecasting.
Yu \etal{} developed a novel temporal matrix factorization model to explicitly represent the network as a function of time \cite{yu2017temporally}. They provided results for link prediction as a specific example and showed that their model performs better than the state-of-the-art techniques.

The most relevant works to this study are available in \cite{hajibagheri2016leveraging, aggarwal2012dynamic, sett2017temporal, amin:ecir19, sun2012will}. The Authors in \cite{hajibagheri2016leveraging} approach the problem of time series link prediction by extracting simple temporal features from the time series, such as mean, (weighted) moving average, and exponential smoothing besides some topological features like common neighbor and Adamic-Adar. But their method is designed for homogeneous networks and fail to consider the heterogeneity of modern networks. Aggarwal \etal{} \cite{aggarwal2012dynamic} tackle the link prediction problem in both dynamic and heterogeneous information networks using a dynamic clustering approach alongside content-based and structural models. However, they aim to solve the conventional link prediction problem, not the continuous-time relationship prediction studied in this paper. In \cite{sett2017temporal}, the authors proposed a feature set, called TMLP, well suited for link prediction in dynamic and heterogeneous information networks. Although their proposed feature set copes with both dynamicity and heterogeneity of the network, it cannot be extended for the generalized problem of relationship prediction and is only designed for solving the simpler link prediction problem. Milani~Fard \etal{} developed an approach called MetaDynaMix which utilizes a set of latent and topological features for predicting a target relationship between two nodes in a dynamic heterogeneous information network \cite{amin:ecir19}. They combine meta path-based topological features with inferred latent features that take temporal network evolutions into account, in order to capture both heterogeneity and dynamicity of the network.

Most of the aforementioned works answered the question of \emph{whether} a link will appear in the network. To the best of our knowledge, the only work that has focused on the continuous-time relationship prediction problem is proposed by Sun \etal{} \cite{sun2012will}, in which a generalized linear model based framework is suggested to model the relationship building time. They consider the building time of links as independent random variables coming from a pre-specified distribution and model the expectation as a function of a linear predictor of the extracted topological features. A shortcoming of this model is that we need to exactly specify the underlying distribution of relationship building times. We came over this problem by learning the distribution from the data using a non-parametric solution. Furthermore, we considered the temporal dynamics of the network which has been entirely ignored in their work.

\section{Conclusion}\label{sec:conclusion}
In this paper, we studied the problem of continuous-time relationship prediction in both dynamic and heterogeneous information networks. To effectively tackle this problem, we first introduced a novel feature extraction framework based on meta-path modeling and recurrent neural network autoencoders to systematically extract features that take both the temporal dynamics and heterogeneous characteristics of the network into account for solving the continuous-time relationship problem. We then proposed a supervised non-parametric model, called \npglm, which exploits the extracted features to predict the relationship building time in information networks. The strength of our model is that it does not impose any significant assumptions on the underlying distribution of the relationship building time given its features, but tries to infer it from the data via a non-parametric approach. Extensive experiments conducted on a synthetic dataset and real-world datasets from DBLP, Delicious, and MovieLens demonstrated the correctness of our method and its effectiveness in predicting the relationship building time.

{For future work, we would like to design a unified architecture to combine feature extraction step with the learning algorithm in an integrated deep learning framework. Moreover, although the propsed method is able to scale to large information networks with thousands of nodes, it is not currently extensible to web-scale information networks where the number of nodes is in the scale of hundreds of millions. Learning temporal non-parametric models within an extremely huge dataset is a challenging problem and is an interesting and important future work. As calculating meta-path-based features are the primary computational bottleneck of our method, to make the learning process scalable, we set to investigate node embedding and approximation techniques.}

\begin{acks}
This work is partially supported by NSF through grant IIS-1763365.
\end{acks}

\bibliographystyle{ACM-Reference-Format}
\bibliography{references}

\end{document}